\documentstyle[epsfig,12pt,axodraw,amssymb]{article}

\newcommand{\sm}[1]{{\scriptscriptstyle #1}}

\def\l{\left(}
\def\r{\right)}

\begin{document}

%%%%%%%%%%%%% Title
\title{Low $\tan{\beta}$ in the extended Minimal Gauge
Mediated Model
\footnote{Talk presented  at the 10th
International Seminar ``Quarks-98'', Suzdal, Russia, May 18-24; to
appear in the Proceedings.}}
\author{  S.~L.~Dubovsky\footnote{E-mail: sergd@ms2.inr.ac.ru},
  D.~S.~Gorbunov\footnote{E-mail: gorby@ms2.inr.ac.ru}\\
  {\small\it
     Institute for Nuclear Research of
         the Russian Academy of Sciences,}\\{\small\it 117312 Moscow, 
Russia
  }}
\date{}
\maketitle
\begin{abstract}

We consider the Minimal Gauge Mediated Model (MGMM) with either fundamental
or antisymmetric messenger multiplets and study consequences of 
mes\-sen\-ger-matter mixing. It is shown that constraints on 
relevant coupling constants
and mixing parameters coming from the existing experimental limits on 
lepton flavor violation and FCNC processes allow 
a wide range of $\tan\beta$
in MGMM with mixing. 
\end{abstract}

\section{Introduction}
In this paper
we discuss a class of SUSY models with gauge mediated supersymmetry
breaking \cite{dine}. In these models supersymmetry is broken 
in a special secluded sector due to  nonperturbative 
dynamics of the corresponding gauge group. 
The soft terms in the
visible sector appear through usual gauge interactions between ordinary
fields and some new set of fields ({\it messengers}) charged
under MSSM gauge group. Messengers obtain supersymmetry breaking
masses through
gauge or Yukawa interactions with the secluded sector. 
In these models gauge interactions transmitting
supersymmetry breaking to the visible sector do not
lead to flavor violation because these interactions are
flavor blind.  Nevertheless, flavor changing interactions can appear
in gauge mediated models due to possible messenger-matter Yukawa
interactions. 

The purpose of this talk is to show that in a reasonable range of
parameters, messenger-matter mixing in Minimal Gauge Mediated Model
(MGMM)\cite{kolda} can give rise to the observable rates of rare
lepton flavor violating processes like $\mu~\to~e\gamma$ and $\mu - e$
conversion, can play a significant role in quark flavor physics and at
the same time can affect radiative electroweak symmetry breaking crucially. 
 
We  find out that, unlike MGMM without mixing, 
at messenger masses of order $10^5$~GeV and
messenger-matter Yukawa
couplings of order $10^{-2}$ (such Yukawas are acceptable 
from the points of view of both theory and 
experiment), there is a wide region
of allowed $\tan{\beta}$ in the model with mixing. The value of
$\tan{\beta}$ is determined by the magnitude of the mixing terms. This
observation leads to an interesting possibility to relate the rates of
flavor violating processes to the Higgs sector parameters.

\section{The model}
\label{s2}

The MGMM contains, in addition to MSSM particles, two messenger
multiplets $Q_{\sm{M}}$ and $\bar{Q}_{\sm{M}}$ which belong to ${\bf
  5}$ and ${\bf \bar{5}}$ representations of $SU(5)$.  We will consider
also the antisymmetric messengers (${\bf 10}$ and ${\bf \bar{10}}$).
All other representations are ruled out by the requirement of
asymptotic freedom of the theory. Messengers couple to a MSSM
singlet $Z$ through the superpotential term
  
$$
{\cal L}_{ms} = \lambda Z Q_{\sm{M}}\bar{Q}_{\sm{M}}.
$$
$Z$ obtains non-vanishing vacuum expectation values $F$ and $S$
via hidden-sector interactions,
$$
Z = S+F\theta \theta. 
$$
Messenger masses and soft breaking terms are expressed most conveniently 
via parameters 
$$
\Lambda=\frac{F}{S}~~~\mbox{and}~~~x=\frac{\lambda F}{\lambda^2S^2}
$$ 
Then masses of fermionic components of messenger superfields
are all equal to $M=\frac{\Lambda}{x}$. 
The vacuum expectation
value of $Z$ mixes scalar components of messenger fields and
gives them masses
$$
M_{\pm}^2=\frac{\Lambda^2}{x^2}(1\pm x)
$$ 
It is clear that $x$ must be smaller than $1$.

\begin{figure}[tb]
\begin{picture}(0,0)%
\epsfig{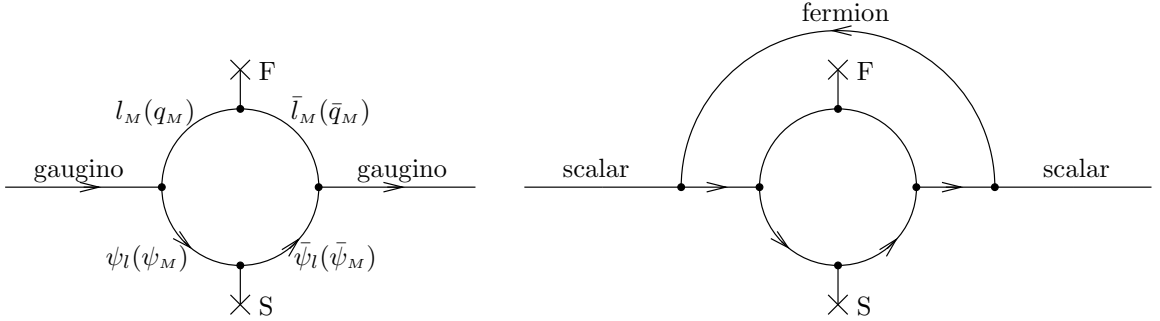}%
\end{picture}%
\setlength{\unitlength}{3947sp}%
\begingroup\makeatletter\ifx\SetFigFont\undefined%
\gdef\SetFigFont#1#2#3#4#5{%
  \reset@font\fontsize{#1}{#2pt}%
  \fontfamily{#3}\fontseries{#4}\fontshape{#5}%
  \selectfont}%
\fi\endgroup%
\begin{picture}(7224,1956)(589,-1527)
\put(2200,-28){\makebox(0,0)[lb]{\smash{\SetFigFont{10}{12.0}{\familydefault}{\mddefault}{\updefault}F}}}
\put(2200,-1505){\makebox(0,0)[lb]{\smash{\SetFigFont{10}{12.0}{\familydefault}{\mddefault}{\updefault}S}}}
\put(5955,-28){\makebox(0,0)[lb]{\smash{\SetFigFont{10}{12.0}{\familydefault}{\mddefault}{\updefault}F}}}
\put(5955,-1505){\makebox(0,0)[lb]{\smash{\SetFigFont{10}{12.0}{\familydefault}{\mddefault}{\updefault}S}}}
\put(786,-643){\makebox(0,0)[lb]{\smash{\SetFigFont{10}{12.0}{\familydefault}{\mddefault}{\updefault}gaugino}}}
\put(2817,-643){\makebox(0,0)[lb]{\smash{\SetFigFont{10}{12.0}{\familydefault}{\mddefault}{\updefault}gaugino}}}
\put(4109,-643){\makebox(0,0)[lb]{\smash{\SetFigFont{10}{12.0}{\familydefault}{\mddefault}{\updefault}scalar}}}
\put(7124,-643){\makebox(0,0)[lb]{\smash{\SetFigFont{10}{12.0}{\familydefault}{\mddefault}{\updefault}scalar}}}
\put(5610,341){\makebox(0,0)[lb]{\smash{\SetFigFont{10}{12.0}{\familydefault}{\mddefault}{\updefault}fermion}}}
\put(1300,-275){\makebox(0,0)[lb]{\smash{\SetFigFont{10}{12.0}{\familydefault}{\mddefault}{\updefault}$l_{\sm{M}}(q_{\sm{M}})$}}}
\put(2400,-275){\makebox(0,0)[lb]{\smash{\SetFigFont{10}{12.0}{\familydefault}{\mddefault}{\updefault}$\bar{l}_{\sm{M}}(\bar{q}_{\sm{M}})$}}}
\put(1240,-1190){\makebox(0,0)[lb]{\smash{\SetFigFont{10}{12.0}{\familydefault}{\mddefault}{\updefault}$\psi_{l}(\psi_{\sm{M}})$}}}
\put(2420,-1190){\makebox(0,0)[lb]{\smash{\SetFigFont{10}{12.0}{\familydefault}{\mddefault}{\updefault}$\bar{\psi}_{l}(\bar{\psi}_{\sm{M}})$}}}
\put(4884,-643){\makebox(0,0)[lb]{\smash{\SetFigFont{10}{12.0}{\familydefault}{\mddefault}{\updefault}}}}
\put(6349,-631){\makebox(0,0)[lb]{\smash{\SetFigFont{10}{12.0}{\familydefault}{\mddefault}{\updefault}}}}
\put(5315,-324){\makebox(0,0)[lb]{\smash{\SetFigFont{10}{12.0}{\familydefault}{\mddefault}{\updefault}}}}
\put(6091,-238){\makebox(0,0)[lb]{\smash{\SetFigFont{10}{12.0}{\familydefault}{\mddefault}{\updefault}}}}
\put(5241,-1099){\makebox(0,0)[lb]{\smash{\SetFigFont{10}{12.0}{\familydefault}{\mddefault}{\updefault}}}}
\put(6300,-1160){\makebox(0,0)[lb]{\smash{\SetFigFont{10}{12.0}{\familydefault}{\mddefault}{\updefault}}}}
\end{picture}
\caption
{ Typical diagrams inducing masses for the MSSM gauginos and scalars. 
Messenger fields run in loops.}
\label{gc}
\end{figure}

Gauginos and scalar particles of MSSM obtain masses at messenger scale
in one and two loops respectively (see Fig.~\ref{gc}). 
Their values are \cite{dine-f}
\begin{equation}
\label{gaugin}
M_{\lambda_{\sm{i}}}(M) = c_i\frac{\alpha_i(M)}{4\pi}\Lambda f_1(x)
\end{equation}
for gauginos and
\begin{equation}
\label{scalmas}
\tilde{m}^2(M)=2\Lambda^2\sum_{i=1}^3c_iC_i\l\frac{\alpha_i(M)}
{4\pi}\r^2f_2(x)
\end{equation}
for scalars. Here $\alpha_i$ are gauge coupling constants of
$SU(3)\times SU(2)\times U(1)$,
$C_i$ are values of the quadratic Casimir operator for 
the pertinent
matter fields: 
$C_3=4/3$ for color triplets (zero for singlets),
 $C_2=3/4$ for weak doublets (zero for singlets),
$C_1=\l\frac{Y}{2}\r^2$, where $Y$ is the weak
 hypercharge. For messengers belonging to the fundamental and antisymmetric
representations
one has $c_1=5/3,~c_2=c_3=1$ and $c_1=5,~c_2=c_3=3$, respectively.   
The dependence of the soft masses on $x$ is very mild, as the functions
$f_1(x)$ and $f_2(x)$ do not deviate much from
1~\cite{Dimopoulos,Martin}. 
The MSSM spectrum at electroweak scale is obtained from
Eqs.~(\ref{gaugin}), (\ref{scalmas}) by making use of renormalizaton group
equations. 

The trilinear soft terms appear in two loops, therefore they are
suppressed by additional gauge coupling constant 
as compared to soft
masses. MGMM contains in addition to $\Lambda$ and $x$ one 
more parameter $\mu$ determining supersymmetric Higgs masses, 
$\mu H_{\sm U}H_{\sm
D}$. Soft mixing $B_\mu h_{\sm U}h_{\sm D}$ appears 
in two loops like other scalar masses, but it is suppressed due to 
accidental cancellations.  

It has been argued in ref.\cite{borzumati} that $\Lambda$ must be
larger than $70$~TeV, otherwise the theory would be
inconsistent with mass limits from LEP.  The characteristic features
of the model without messenger-matter mixing 
are large $\tan{\beta}$~\cite{kolda} (an estimate of
refs.\cite{1.3,borzumati} is $\tan \beta\gtrsim 50$) and large squark masses.
The parameter $\mu$ is predicted to be about $500$ GeV. 
There is large mixing between $\tilde{\tau}_{{\sm R}}$ and
$\tilde{\tau}_{{\sm L}}$, proportional to $\tan{\beta}$ and $\mu$.
It results in mass splitting of $\tau$-sleptons so that 
the Next Lightest Supersymmetric
Particle (NLSP) is a combination of $\tilde{\tau}_{{\sm R}}$ and
$\tilde{\tau}_{{\sm L}}$ \cite{kolda,borzumati}, the LSP being
gravitino. 

\section{Messenger-matter mixing}

Messenger fields may be odd or even under R-parity. 
Fundamental messengers, depending on their R-parity, 
have the same quantum numbers as either
fundamental matter or fundamental Higgs fields,
so messenger-matter mixing arises naturally. In the latter case
triplet messenger fields will give rise to fast proton decay due
to possible Higgs-like mixing with ordinary fields, unless the
corresponding Yukawa couplings are smaller than
$10^{-21}$~\cite{R-even}. 

We will consider messengers which are odd under
R-parity. Then the components of the fundamental messengers
$Q_{\sm{M}}^{(5)}=(l_{\sm{M}},q_{\sm{M}})$ have the same quantum
numbers as left leptons and  right down-quarks, while components of
antisymmetric messengers $Q_{\sm{M}}^{(10)}$ have quantum numbers of
the right leptons, left quarks and right
 up-quarks.
We assume that there is one generation of messengers, fundamental 
or antisymmetric, and consider their mixing with
ordinary matter separately.

One can introduce messenger-matter mixing~\cite{dinem}
\begin{equation}
\label{y-term}
{\cal W}_{mm}^{(5)} = H_{\sm D}L_{\hat{i}}Y^{(5)}_{\hat{i}j}E_j+H_{\sm D}
D_{\hat{i}}X^{(5)}_{\hat{i}j}Q_j
\end{equation}
in the case of fundamental representation and 

\begin{equation}
\label{y-term-10}
{\cal W}_{mm}^{(10)} = H_{\sm D}E_{\hat{i}}Y^{(10)}_{\hat{i}j}L_j+H_{\sm D}
Q_{\hat{i}}W^{(10)}_{\hat{i}j}D_j +
H_{\sm U}
U_{\hat{i}}X^{(10)}_{\hat{i}\hat{j}}Q_{\hat{j}}
\end{equation}
in the case of antisymmetric messenger fields, where we use the notation 
\begin{displaymath}
L_{\hat{i}}=\l { e_{\sm L} \choose \nu_e },{\mu_{\sm L} \choose \nu_\mu},
{\tau_{\sm L} \choose \nu_\tau},{l_{m_{\sm L}} \choose \nu_{m_{\sm L}}}\r
\end{displaymath}
for the left doublet superfields in the lepton sector 
and similarly for the quark sector.

Hereafter $\hat{i},\hat{j}=1,..,4$ label the three matter 
generations and the messenger field, $i,j=1,..,3$
correspond to the three matter generations. 
$Y^{(5)}_{\hat{i}j},X^{(5)}_{\hat{i}j}$,
$Y^{(10)}_{\hat{i}j},W^{(10)}_{\hat{i}j}$  are the
$4\times3$ matrices of Yukawa couplings and 
$X^{(10)}_{\hat{i}\hat{j}}$ is $4\times 4$ matrix. Explicitly 
$$
Y_{\hat{i}j}^{(5)} = 
\left(
\begin{array}{ccc}
Y_{e}& 0 & 0\\
0 & Y_{\mu} & 0\\
0 & 0 & Y_{\tau}\\
Y_{41}^{(5)}& Y_{42}^{(5)}& Y_{43}^{(5)}
\end{array}
\right)
$$
and the other matrices have similar form. 

There is no CP-violation in this
theory in the lepton sector. 
Arbitrary phases may be rotated away by 
redefinition of the lepton fields. On the other hand, there appears 
CP-violation in the quark sector, in  addition to the CKM mechanism. 

To summarize, messenger-matter mixing in the leptonic sector occurs
through the Yukawa couplings $Y_{4i}^{(5)}$ or  $Y_{4i}^{(10)}$
($i=1,2,3$), depending on the representation of the messenger fields.
Likewise, mixing in the quark sector appears through 
$X_{4i}^{(5)}$ or
 $X_{i4}^{(10)}$,
 $X_{4i}^{(10)}$,
 $W_{4i}^{(10)}$. In the following sections we
sometimes use the collective notation $Y_i$ for the 
 couplings 
$Y_{4i}$, $X_{4i}$, $X_{i4}$ and $W_{4i}$ in 
statements applicable to all of them.

One of the most important effects of this mixing is the
absence of heavy stable charged (and colored) particles (messengers)
in the theory~\cite{dinem}.  Other known possibilities to solve this 
problem~\cite{Dimopoulos} faced the necessity of 
fine-tuning of the messenger mass parameters.

The insertion of the messenger-matter mixing terms has interesting
experimental consequences. 
Supersymmetry breaking in MGMM does not lead to flavor violation
because the gauge interactions creating soft terms
in the visible sector are flavor blind. In the
messenger-matter mixing
extension of MGMM we obtain a rich pattern of flavor physics. It is
{\it a priori} clear that the mixing terms induce 
lepton flavor violation as
well as FCNC processes. Moreover, these terms contain new sources of
CP-violation in the quark sector. 

\section{Induced mixing of matter fields}

Let us consider mixing between the fields of MSSM that appears 
after messengers are integrated out.
It is straightforward to check that the fermion mixing terms are small 
at the tree level. In principle, this mixing may lead to lepton and quark
flavor violation due to one loop diagrams involving scalars and
gauginos~\cite{barbieri}.  However, this effect is 
negligible due to see-saw type mechanism.

The main contribution to the processes in the visible sector comes from
slepton and squark mixing. 
Tree level mixing in the scalar sector is 
again small. 
The dominant contributions to mixing in the slepton and the squark sectors 
appear through one loop diagrams originating from
trilinear terms in the
superpotential that involve messengers, 
$H_{\sm D}$ for fundamental messengers,
and also $H_{\sm U}$ for antisymmetric ones.
\footnote{In the case of very small values of $x$
the dominant contributions to $\delta m_{ij}^2$ come from two loops
rather than one loop, but we will not consider  this special
case
(see discussion in
ref.~\cite{b-tan}).}  

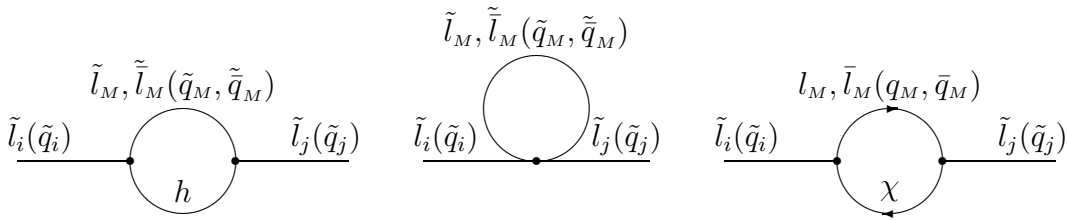
\begin{figure}[htb]
\begin{center}
\unitlength=1.00mm
\special{em:linewidth 0.4pt}
\linethickness{0.4pt}
\begin{picture}(148.00,40.00)(12,100)
\put(10.00,120.00){\line(1,0){15.00}}
\put(32.00,120.00){\circle{14.00}}
\put(39.00,120.00){\line(1,0){15.00}}
\put(64.00,120.00){\line(1,0){30.00}}
\put(79.00,127.00){\circle{14.00}}
\put(13.00,122.00){\makebox(0,0)[cb]{$\tilde{l}_i(\tilde{q}_i)$}}
\put(51.00,122.00){\makebox(0,0)[cb]{$\tilde{l}_j(\tilde{q}_j)$}}
\put(67.00,122.00){\makebox(0,0)[cb]{$\tilde{l}_i(\tilde{q}_i)$}}
\put(91.00,122.00){\makebox(0,0)[cb]{$\tilde{l}_j(\tilde{q}_j)$}}
\put(32.00,129.00){\makebox(0,0)[cb]{$\tilde{l}_{\sm
M},\tilde{\bar{l}}_{\sm M}
(\tilde{q}_{\sm M},\tilde{\bar{q}}_{\sm M})$}}
\put(32.00,116.00){\makebox(0,0)[cc]{$h$}}
\put(79.00,136.00){\makebox(0,0)[cb]{$\tilde{l}_{\sm
M},\tilde{\bar{l}}_{\sm M}
(\tilde{q}_{\sm M},\tilde{\bar{q}}_{\sm M})$}}
\put(104.00,120.00){\line(1,0){15.00}}
\put(126.00,120.00){\circle{14.00}}
\put(133.00,120.00){\line(1,0){15.00}}
\put(107.00,122.00){\makebox(0,0)[cb]{$\tilde{l}_i(\tilde{q}_i)$}}
\put(145.00,122.00){\makebox(0,0)[cb]{$\tilde{l}_j(\tilde{q}_j)$}}
\put(126.00,129.00){\makebox(0,0)[cb]{$l_{\sm M},\bar{l}_{\sm M}(q_{\sm M},\bar{q}_{\sm M})$}}
\put(126.00,116.00){\makebox(0,0)[cc]{$\chi$}}
\put(126.00,127.00){\vector(1,0){1.00}}
\put(126.00,113.00){\vector(-1,0){1.00}}
\put(79.00,120.00){\circle*{1.00}}
\put(39.00,120.00){\circle*{1.00}}
\put(25.00,120.00){\circle*{1.00}}
\put(119.00,120.00){\circle*{1.00}}
\put(133.00,120.00){\circle*{1.00}}
\end{picture} 
\caption{The diagrams dominating the scalar mixing matrix.}
\label{gopa}
\end{center}
\end{figure}

After diagonalizing the messenger mass matrix we obtain the
diagrams contributing to scalar mixing to the order $(\lambda
S)^2$, which are shown in Fig.~\ref{gopa}.
If supersymmetry were unbroken, the sum of these diagrams would be equal to
zero.
In our case of broken supersymmetry the resulting
contributions to the mass matrices of MSSM scalars (for $x$ not 
very close to 1) are
\begin{equation}
\delta m^2_{ij}=-a\frac{\Lambda^2}{96\pi^2}Y^*_iY_jx^2
\label{delta-mixing}
\end{equation}
where $a=1(2)$ for scalars transforming as weak doublets (singlets).  

\section{Constraints on Yukawa matrices}
There are two types of constraints on Yukawa couplings inherent in this
theory.   
\begin{itemize}
\item Since scalars get negative shifts in squared masses, 
the expression~(\ref{delta-mixing}) for the
soft terms immediately implies theoretical bounds on Yukawa
couplings $Y_i$ which come from the requirement \cite{dinem}
that none of the scalar squared masses becomes negative.
\item Mixing~(\ref{y-term}), (\ref{y-term-10}) 
leads to the FCNC processes as well as to lepton
flavor and CP violation. The existing experimental bounds on these 
processes provide  constraints on the products of Yukawa
couplings. 
\end{itemize}

\subsection{Theoretical bounds}
Let us first obtain theoretical constraints on Yukawa
couplings present in this model. With loop 
corrections~(\ref{delta-mixing})  
to the scalar mass matrix included, its
eigenvalues $\tilde{m}_i^2$ are all positive only if
\begin{equation}
\tilde{m}^2+\sum_{j=1}^{3}\delta m^2_{jj}>0
\label{delta}
\end{equation}
where 
$\tilde{m}^2$ is given by eq.~(\ref{scalmas}).
Making use of inequality $|Y_{i}Y_{j}|<\frac{1}{2}\sum{|Y_{i}|^2}$ 
one obtains from eq.~(\ref{delta})
the theoretical constraint on mixing 
\begin{equation}
-\delta\tilde{m}^2_{ij}<\frac{1}{2} \tilde{m}^2
\label{mix_c}
\end{equation} 
As an example, the self-consistency 
condition of the theory can be written in case of small $x$ as follows,
\begin{equation}
Y_{i}Y_{j}x^2<\frac{96\pi^2}{a}
\sum_{i=1}^{3}c_iC_i\l\frac{\alpha_i}{4\pi}\r^2
\label{sc_2}
\end{equation}

\subsection{The experimental limits}

Let us now consider the effects of scalar mixing on the usual leptons
and quarks.  
It is clear that different experiments are
sensitive, generally speaking, to different combinations of Yukawa 
couplings and CP-violating phases. 

The bounds on  various products 
of Yukawa couplings between messengers and quarks (and leptons) 
coming from the
requirement that corresponding mixing is consistent with the
present experimental limits
on the rare processes were found in refs.~\cite{we,b-tan}. 
Some of the theoretical bounds 
coming from eq.~(\ref{sc_2}) are stronger than corresponding 
experimental ones.

The typical constraint on Yukawa couplings in the quark sector is
$Yx\lesssim 0.1$ and comes from mixing of neutral mesons, 
while $b\to s\gamma$ decay does not give any bounds and the 
corresponding Yukawa terms are limited by the self-consistency
condition~(\ref{sc_2}). There are also limits on CP-violating terms 
coming from the analysis of $K^0-\bar{K}^0$ system and $K\to\pi\pi$ decays.

In the slepton sector there is a single experimental
constraint coming from the $\mu\to
e\gamma$ decay: 
$|Y^{(5,10)}_{41}Y^{(5,10)}_{42}|^{1/2}x\lesssim
3\times 10^{-2}$.
Slepton mixing in the gauge-mediated models 
leads also to $\mu\to e$ conversion, but 
the existing limit on this process gives weaker 
bound on the product of Yukawa couplings than 
one 
coming from the $\mu\to e\gamma$ decay. Concerning the other processes, the 
self-consistence of the model requires that 
the rates
$\Gamma(\tau \to e\gamma)$ and $\Gamma(\tau \to \mu\gamma)$
 are lower
than the present experimental limits at least by factor of 
$10^{-2}$. The same conclusion holds in the case of 
antisymmetric messengers as well.

Therefore we may conclude that sizeable messenger-matter 
mixing is consistent both with experimental limits and with
electroweak symmetry breaking pattern.

\section{Electroweak breaking and squark masses}
It was already mentioned  that radiative electroweak breaking in
MGMM without messenger-matter mixing leads to  large values of $\tan{\beta}\gtrsim 50$.
  A usual way to avoid this limit is to
assume some extra soft contribution to the Higgs sector of the theory.
In this section we consider electroweak breaking in the model with 
messenger-matter mixing. In particular, we show that wide range 
of values of 
$\tan{\beta}$ is now allowed without any 
introducing additional parameters in the Higgs sector
of the model. 

Minimization of the Higgs potential 
\begin{eqnarray}
\nonumber
V=(\mu^2+m^2_{h_{\sm U}})h_{\sm U}^2+(\mu^2+m^2_{h_{\sm D}})h_{\sm
D}^2+B_\mu (h_{\sm U}h_{\sm D}+h.c.)+ \\
\nonumber
\frac{g'^2+g^2}{8}(|h_{\sm U}|^2-|h_{\sm D}|^2)^2+
\frac{g^2}{2}|h_{\sm U}^+h_{\sm D}|^2
\end{eqnarray}
results in the following two
equations, 
\begin{equation}
\label{ewb}
\sin{2\beta}=\frac{-2B_\mu}{m^2_{h_{\sm U}}+m^2_{h_{\sm D}}+2\mu^2}\;,~~~
\mu^2=\frac{m^2_{h_{\sm D}}-
m^2_{h_{\sm U}}\tan^2{\beta}}{\tan^2{\beta}-1}-\frac{1}{2}M_Z^2
\end{equation}
At the two loop level, the
parameter $B_\mu$ characterizing the magnitude of the soft 
mixing term $B_\mu h_{\sm U}h_{\sm D}$ 
in the Higgs sector is equal to~\cite{kolda}
\begin{equation}
\label{B-term}
B_\mu=M_{\lambda_2}\mu(-0.12+0.17Y_t^2)\;, 
\end{equation}
where $M_{\lambda_2}$ is given by eq.~(\ref{gaugin}). 

In MGMM without messenger-matter mixing,
the value of the soft mass $m^2_{h_{\sm D}}$ is given by 
eq.~(\ref{scalmas}) while $m^2_{h_{\sm U}}$ receives additional large negative 
one-loop correction due to large Yukawa coupling between $H_{\sm U}$ and top-quark,
\begin{equation}
\label{toploop}
\delta m^2_{h_{\sm U}}=-\frac{3Y^2_t}{4\pi^2}m^2_{\tilde{t}}\ln\left(\frac
{\Lambda}{xm_{\tilde{t}}}\right)
\end{equation}
It follows from
eqs.~(\ref{scalmas}) and (\ref{toploop}) that 
$\delta m^2_{h_{\sm U}}\gtrsim m^2_{h_{\sm U}},m^2_{h_{\sm D}}$, hence
eq.~(\ref{ewb}) takes the following simple form
\begin{equation}
\mu^2\simeq -\delta m^2_{h_{\sm U}}\;,~~~~\sin{2\beta}\simeq -
\frac{2B_\mu}{\mu^2}
\label{large}
\end{equation}
It is clear from eqs.~(\ref{B-term}) and (\ref{large}) that $\sin 2\beta\ll 1$,
and, therefore, $\tan\beta\gg 1$. This simple estimate gives
$\tan{\beta}\geq 50$. Detailed calculations~\cite{1.3,borzumati} confirm this 
estimate in MGMM without mixing.

In the presence of messenger-matter 
mixing squared Higgs masses receive negative  
contribution from the diagrams 
analogous to those shown in Fig.~\ref{gopa}: 
\begin{equation}
\delta m^2_{h_{\sm D}}=-d_{\sm D}\frac{\Lambda^2}{96\pi^2}x^2\;,~~~
\delta m^2_{h_{\sm U}}=-d_{\sm U}\frac{\Lambda^2}{96\pi^2}x^2
\label{dloop}
\end{equation}
where 
$$
d_{\sm
D}=\sum_{i=1}^3\left(|Y_{4i}^{(5)}|^2+|Y_{4i}^{(10)}|^2+3|W_{4i}^{(10)}|^2+
3|X_{4i}^{(5)}|^2\right),
$$$$
d_{\sm
U}=\sum_{i=1}^3\left(3|X_{4i}^{(10)}|^2+3|X_{i4}^{(10)}|^2\right)
$$
and we consider ``small'' $x$ ($x\lesssim 0.8$). 

Depending on the Yukawa coupling constants, the mass splitting 
$\delta m^2_{h_{\sm D}}$ may be of the same order as 
$\delta m^2_{h_{\sm U}}$. In that case the value of $\sin{2\beta}$ 
gets modified as compared to eq.~(\ref{large}). One has an estimate 
\begin{equation}
\mu^2\simeq\delta m^2_{h_{\sm U}},~~~~\sin 2\beta\simeq 
\frac{-2B_\mu}{\mu^2-\delta m^2_{h_{\sm D}}}
\label{small}
\end{equation}
Therefore messenger-matter mixing reduces $\tan\beta$. 
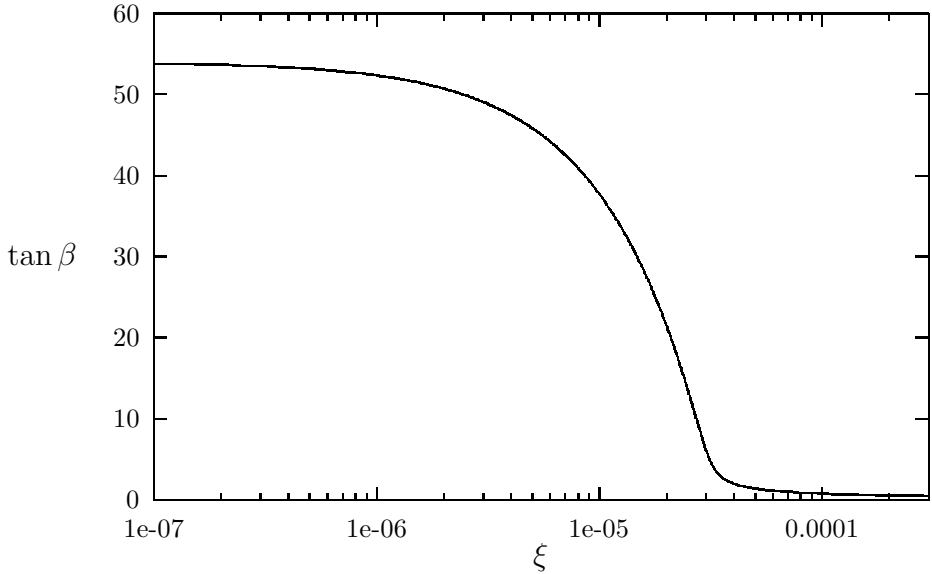
\begin{figure}[htb]
% GNUPLOT: LaTeX picture
\setlength{\unitlength}{0.240900pt}
\ifx\plotpoint\undefined\newsavebox{\plotpoint}\fi
\begin{picture}(1500,900)(0,0)
\font\gnuplot=cmr10 at 10pt
\gnuplot
\sbox{\plotpoint}{\rule[-0.200pt]{0.400pt}{0.400pt}}%
\put(220.0,113.0){\rule[-0.200pt]{292.934pt}{0.400pt}}
\put(220.0,113.0){\rule[-0.200pt]{4.818pt}{0.400pt}}
\put(198,113){\makebox(0,0)[r]{0}}
\put(1416.0,113.0){\rule[-0.200pt]{4.818pt}{0.400pt}}
\put(220.0,240.0){\rule[-0.200pt]{4.818pt}{0.400pt}}
\put(198,240){\makebox(0,0)[r]{10}}
\put(1416.0,240.0){\rule[-0.200pt]{4.818pt}{0.400pt}}
\put(220.0,368.0){\rule[-0.200pt]{4.818pt}{0.400pt}}
\put(198,368){\makebox(0,0)[r]{20}}
\put(1416.0,368.0){\rule[-0.200pt]{4.818pt}{0.400pt}}
\put(220.0,495.0){\rule[-0.200pt]{4.818pt}{0.400pt}}
\put(198,495){\makebox(0,0)[r]{30}}
\put(1416.0,495.0){\rule[-0.200pt]{4.818pt}{0.400pt}}
\put(220.0,622.0){\rule[-0.200pt]{4.818pt}{0.400pt}}
\put(198,622){\makebox(0,0)[r]{40}}
\put(1416.0,622.0){\rule[-0.200pt]{4.818pt}{0.400pt}}
\put(220.0,750.0){\rule[-0.200pt]{4.818pt}{0.400pt}}
\put(198,750){\makebox(0,0)[r]{50}}
\put(1416.0,750.0){\rule[-0.200pt]{4.818pt}{0.400pt}}
\put(220.0,877.0){\rule[-0.200pt]{4.818pt}{0.400pt}}
\put(198,877){\makebox(0,0)[r]{60}}
\put(1416.0,877.0){\rule[-0.200pt]{4.818pt}{0.400pt}}
\put(220.0,113.0){\rule[-0.200pt]{0.400pt}{4.818pt}}
\put(220,68){\makebox(0,0){1e-07}}
\put(220.0,857.0){\rule[-0.200pt]{0.400pt}{4.818pt}}
\put(325.0,113.0){\rule[-0.200pt]{0.400pt}{2.409pt}}
\put(325.0,867.0){\rule[-0.200pt]{0.400pt}{2.409pt}}
\put(387.0,113.0){\rule[-0.200pt]{0.400pt}{2.409pt}}
\put(387.0,867.0){\rule[-0.200pt]{0.400pt}{2.409pt}}
\put(431.0,113.0){\rule[-0.200pt]{0.400pt}{2.409pt}}
\put(431.0,867.0){\rule[-0.200pt]{0.400pt}{2.409pt}}
\put(464.0,113.0){\rule[-0.200pt]{0.400pt}{2.409pt}}
\put(464.0,867.0){\rule[-0.200pt]{0.400pt}{2.409pt}}
\put(492.0,113.0){\rule[-0.200pt]{0.400pt}{2.409pt}}
\put(492.0,867.0){\rule[-0.200pt]{0.400pt}{2.409pt}}
\put(516.0,113.0){\rule[-0.200pt]{0.400pt}{2.409pt}}
\put(516.0,867.0){\rule[-0.200pt]{0.400pt}{2.409pt}}
\put(536.0,113.0){\rule[-0.200pt]{0.400pt}{2.409pt}}
\put(536.0,867.0){\rule[-0.200pt]{0.400pt}{2.409pt}}
\put(554.0,113.0){\rule[-0.200pt]{0.400pt}{2.409pt}}
\put(554.0,867.0){\rule[-0.200pt]{0.400pt}{2.409pt}}
\put(570.0,113.0){\rule[-0.200pt]{0.400pt}{4.818pt}}
\put(570,68){\makebox(0,0){1e-06}}
\put(570.0,857.0){\rule[-0.200pt]{0.400pt}{4.818pt}}
\put(675.0,113.0){\rule[-0.200pt]{0.400pt}{2.409pt}}
\put(675.0,867.0){\rule[-0.200pt]{0.400pt}{2.409pt}}
\put(737.0,113.0){\rule[-0.200pt]{0.400pt}{2.409pt}}
\put(737.0,867.0){\rule[-0.200pt]{0.400pt}{2.409pt}}
\put(780.0,113.0){\rule[-0.200pt]{0.400pt}{2.409pt}}
\put(780.0,867.0){\rule[-0.200pt]{0.400pt}{2.409pt}}
\put(814.0,113.0){\rule[-0.200pt]{0.400pt}{2.409pt}}
\put(814.0,867.0){\rule[-0.200pt]{0.400pt}{2.409pt}}
\put(842.0,113.0){\rule[-0.200pt]{0.400pt}{2.409pt}}
\put(842.0,867.0){\rule[-0.200pt]{0.400pt}{2.409pt}}
\put(865.0,113.0){\rule[-0.200pt]{0.400pt}{2.409pt}}
\put(865.0,867.0){\rule[-0.200pt]{0.400pt}{2.409pt}}
\put(886.0,113.0){\rule[-0.200pt]{0.400pt}{2.409pt}}
\put(886.0,867.0){\rule[-0.200pt]{0.400pt}{2.409pt}}
\put(903.0,113.0){\rule[-0.200pt]{0.400pt}{2.409pt}}
\put(903.0,867.0){\rule[-0.200pt]{0.400pt}{2.409pt}}
\put(919.0,113.0){\rule[-0.200pt]{0.400pt}{4.818pt}}
\put(919,68){\makebox(0,0){1e-05}}
\put(919.0,857.0){\rule[-0.200pt]{0.400pt}{4.818pt}}
\put(1025.0,113.0){\rule[-0.200pt]{0.400pt}{2.409pt}}
\put(1025.0,867.0){\rule[-0.200pt]{0.400pt}{2.409pt}}
\put(1086.0,113.0){\rule[-0.200pt]{0.400pt}{2.409pt}}
\put(1086.0,867.0){\rule[-0.200pt]{0.400pt}{2.409pt}}
\put(1130.0,113.0){\rule[-0.200pt]{0.400pt}{2.409pt}}
\put(1130.0,867.0){\rule[-0.200pt]{0.400pt}{2.409pt}}
\put(1164.0,113.0){\rule[-0.200pt]{0.400pt}{2.409pt}}
\put(1164.0,867.0){\rule[-0.200pt]{0.400pt}{2.409pt}}
\put(1192.0,113.0){\rule[-0.200pt]{0.400pt}{2.409pt}}
\put(1192.0,867.0){\rule[-0.200pt]{0.400pt}{2.409pt}}
\put(1215.0,113.0){\rule[-0.200pt]{0.400pt}{2.409pt}}
\put(1215.0,867.0){\rule[-0.200pt]{0.400pt}{2.409pt}}
\put(1235.0,113.0){\rule[-0.200pt]{0.400pt}{2.409pt}}
\put(1235.0,867.0){\rule[-0.200pt]{0.400pt}{2.409pt}}
\put(1253.0,113.0){\rule[-0.200pt]{0.400pt}{2.409pt}}
\put(1253.0,867.0){\rule[-0.200pt]{0.400pt}{2.409pt}}
\put(1269.0,113.0){\rule[-0.200pt]{0.400pt}{4.818pt}}
\put(1269,68){\makebox(0,0){0.0001}}
\put(1269.0,857.0){\rule[-0.200pt]{0.400pt}{4.818pt}}
\put(1374.0,113.0){\rule[-0.200pt]{0.400pt}{2.409pt}}
\put(1374.0,867.0){\rule[-0.200pt]{0.400pt}{2.409pt}}
\put(1436.0,113.0){\rule[-0.200pt]{0.400pt}{2.409pt}}
\put(1436.0,867.0){\rule[-0.200pt]{0.400pt}{2.409pt}}
\put(220.0,113.0){\rule[-0.200pt]{292.934pt}{0.400pt}}
\put(1436.0,113.0){\rule[-0.200pt]{0.400pt}{184.048pt}}
\put(220.0,877.0){\rule[-0.200pt]{292.934pt}{0.400pt}}
\put(45,495){\makebox(0,0){$\tan{\beta}$}}
\put(828,23){\makebox(0,0){$\xi$}}
\put(220.0,113.0){\rule[-0.200pt]{0.400pt}{184.048pt}}
\put(1353,118.67){\rule{18.308pt}{0.400pt}}
\multiput(1391.00,118.17)(-38.000,1.000){2}{\rule{9.154pt}{0.400pt}}
\put(1301,119.67){\rule{12.527pt}{0.400pt}}
\multiput(1327.00,119.17)(-26.000,1.000){2}{\rule{6.263pt}{0.400pt}}
\put(1263,121.17){\rule{7.700pt}{0.400pt}}
\multiput(1285.02,120.17)(-22.018,2.000){2}{\rule{3.850pt}{0.400pt}}
\put(1235,122.67){\rule{6.745pt}{0.400pt}}
\multiput(1249.00,122.17)(-14.000,1.000){2}{\rule{3.373pt}{0.400pt}}
\put(1213,124.17){\rule{4.500pt}{0.400pt}}
\multiput(1225.66,123.17)(-12.660,2.000){2}{\rule{2.250pt}{0.400pt}}
\put(1195,125.67){\rule{4.336pt}{0.400pt}}
\multiput(1204.00,125.17)(-9.000,1.000){2}{\rule{2.168pt}{0.400pt}}
\put(1181,126.67){\rule{3.373pt}{0.400pt}}
\multiput(1188.00,126.17)(-7.000,1.000){2}{\rule{1.686pt}{0.400pt}}
\put(1170,128.17){\rule{2.300pt}{0.400pt}}
\multiput(1176.23,127.17)(-6.226,2.000){2}{\rule{1.150pt}{0.400pt}}
\put(1160,129.67){\rule{2.409pt}{0.400pt}}
\multiput(1165.00,129.17)(-5.000,1.000){2}{\rule{1.204pt}{0.400pt}}
\put(1152,131.17){\rule{1.700pt}{0.400pt}}
\multiput(1156.47,130.17)(-4.472,2.000){2}{\rule{0.850pt}{0.400pt}}
\put(1146,132.67){\rule{1.445pt}{0.400pt}}
\multiput(1149.00,132.17)(-3.000,1.000){2}{\rule{0.723pt}{0.400pt}}
\put(1140,133.67){\rule{1.445pt}{0.400pt}}
\multiput(1143.00,133.17)(-3.000,1.000){2}{\rule{0.723pt}{0.400pt}}
\put(1135,135.17){\rule{1.100pt}{0.400pt}}
\multiput(1137.72,134.17)(-2.717,2.000){2}{\rule{0.550pt}{0.400pt}}
\put(1131,136.67){\rule{0.964pt}{0.400pt}}
\multiput(1133.00,136.17)(-2.000,1.000){2}{\rule{0.482pt}{0.400pt}}
\put(1127,138.17){\rule{0.900pt}{0.400pt}}
\multiput(1129.13,137.17)(-2.132,2.000){2}{\rule{0.450pt}{0.400pt}}
\put(1124,139.67){\rule{0.723pt}{0.400pt}}
\multiput(1125.50,139.17)(-1.500,1.000){2}{\rule{0.361pt}{0.400pt}}
\put(1121,140.67){\rule{0.723pt}{0.400pt}}
\multiput(1122.50,140.17)(-1.500,1.000){2}{\rule{0.361pt}{0.400pt}}
\put(1119,142.17){\rule{0.482pt}{0.400pt}}
\multiput(1120.00,141.17)(-1.000,2.000){2}{\rule{0.241pt}{0.400pt}}
\put(1116,143.67){\rule{0.723pt}{0.400pt}}
\multiput(1117.50,143.17)(-1.500,1.000){2}{\rule{0.361pt}{0.400pt}}
\put(1114,145.17){\rule{0.482pt}{0.400pt}}
\multiput(1115.00,144.17)(-1.000,2.000){2}{\rule{0.241pt}{0.400pt}}
\put(1112,146.67){\rule{0.482pt}{0.400pt}}
\multiput(1113.00,146.17)(-1.000,1.000){2}{\rule{0.241pt}{0.400pt}}
\put(1111,147.67){\rule{0.241pt}{0.400pt}}
\multiput(1111.50,147.17)(-0.500,1.000){2}{\rule{0.120pt}{0.400pt}}
\put(1109,149.17){\rule{0.482pt}{0.400pt}}
\multiput(1110.00,148.17)(-1.000,2.000){2}{\rule{0.241pt}{0.400pt}}
\put(1108,150.67){\rule{0.241pt}{0.400pt}}
\multiput(1108.50,150.17)(-0.500,1.000){2}{\rule{0.120pt}{0.400pt}}
\put(1106,152.17){\rule{0.482pt}{0.400pt}}
\multiput(1107.00,151.17)(-1.000,2.000){2}{\rule{0.241pt}{0.400pt}}
\put(1105,153.67){\rule{0.241pt}{0.400pt}}
\multiput(1105.50,153.17)(-0.500,1.000){2}{\rule{0.120pt}{0.400pt}}
\put(1104,154.67){\rule{0.241pt}{0.400pt}}
\multiput(1104.50,154.17)(-0.500,1.000){2}{\rule{0.120pt}{0.400pt}}
\put(1102.67,156){\rule{0.400pt}{0.482pt}}
\multiput(1103.17,156.00)(-1.000,1.000){2}{\rule{0.400pt}{0.241pt}}
\put(1102,157.67){\rule{0.241pt}{0.400pt}}
\multiput(1102.50,157.17)(-0.500,1.000){2}{\rule{0.120pt}{0.400pt}}
\put(1100.67,159){\rule{0.400pt}{0.482pt}}
\multiput(1101.17,159.00)(-1.000,1.000){2}{\rule{0.400pt}{0.241pt}}
\put(1100,160.67){\rule{0.241pt}{0.400pt}}
\multiput(1100.50,160.17)(-0.500,1.000){2}{\rule{0.120pt}{0.400pt}}
\put(1099,161.67){\rule{0.241pt}{0.400pt}}
\multiput(1099.50,161.17)(-0.500,1.000){2}{\rule{0.120pt}{0.400pt}}
\put(1097.67,163){\rule{0.400pt}{0.482pt}}
\multiput(1098.17,163.00)(-1.000,1.000){2}{\rule{0.400pt}{0.241pt}}
\put(1097,164.67){\rule{0.241pt}{0.400pt}}
\multiput(1097.50,164.17)(-0.500,1.000){2}{\rule{0.120pt}{0.400pt}}
\put(1429.0,119.0){\rule[-0.200pt]{1.686pt}{0.400pt}}
\put(1096,167.67){\rule{0.241pt}{0.400pt}}
\multiput(1096.50,167.17)(-0.500,1.000){2}{\rule{0.120pt}{0.400pt}}
\put(1095,168.67){\rule{0.241pt}{0.400pt}}
\multiput(1095.50,168.17)(-0.500,1.000){2}{\rule{0.120pt}{0.400pt}}
\put(1097.0,166.0){\rule[-0.200pt]{0.400pt}{0.482pt}}
\put(1094,171.67){\rule{0.241pt}{0.400pt}}
\multiput(1094.50,171.17)(-0.500,1.000){2}{\rule{0.120pt}{0.400pt}}
\put(1092.67,173){\rule{0.400pt}{0.482pt}}
\multiput(1093.17,173.00)(-1.000,1.000){2}{\rule{0.400pt}{0.241pt}}
\put(1095.0,170.0){\rule[-0.200pt]{0.400pt}{0.482pt}}
\put(1092,175.67){\rule{0.241pt}{0.400pt}}
\multiput(1092.50,175.17)(-0.500,1.000){2}{\rule{0.120pt}{0.400pt}}
\put(1093.0,175.0){\usebox{\plotpoint}}
\put(1091,178.67){\rule{0.241pt}{0.400pt}}
\multiput(1091.50,178.17)(-0.500,1.000){2}{\rule{0.120pt}{0.400pt}}
\put(1089.67,180){\rule{0.400pt}{0.482pt}}
\multiput(1090.17,180.00)(-1.000,1.000){2}{\rule{0.400pt}{0.241pt}}
\put(1092.0,177.0){\rule[-0.200pt]{0.400pt}{0.482pt}}
\put(1089,182.67){\rule{0.241pt}{0.400pt}}
\multiput(1089.50,182.17)(-0.500,1.000){2}{\rule{0.120pt}{0.400pt}}
\put(1090.0,182.0){\usebox{\plotpoint}}
\put(1088,185.67){\rule{0.241pt}{0.400pt}}
\multiput(1088.50,185.17)(-0.500,1.000){2}{\rule{0.120pt}{0.400pt}}
\put(1089.0,184.0){\rule[-0.200pt]{0.400pt}{0.482pt}}
\put(1087,188.67){\rule{0.241pt}{0.400pt}}
\multiput(1087.50,188.17)(-0.500,1.000){2}{\rule{0.120pt}{0.400pt}}
\put(1088.0,187.0){\rule[-0.200pt]{0.400pt}{0.482pt}}
\put(1085.67,191){\rule{0.400pt}{0.482pt}}
\multiput(1086.17,191.00)(-1.000,1.000){2}{\rule{0.400pt}{0.241pt}}
\put(1087.0,190.0){\usebox{\plotpoint}}
\put(1084.67,194){\rule{0.400pt}{0.482pt}}
\multiput(1085.17,194.00)(-1.000,1.000){2}{\rule{0.400pt}{0.241pt}}
\put(1086.0,193.0){\usebox{\plotpoint}}
\put(1084,196.67){\rule{0.241pt}{0.400pt}}
\multiput(1084.50,196.17)(-0.500,1.000){2}{\rule{0.120pt}{0.400pt}}
\put(1085.0,196.0){\usebox{\plotpoint}}
\put(1082.67,201){\rule{0.400pt}{0.482pt}}
\multiput(1083.17,201.00)(-1.000,1.000){2}{\rule{0.400pt}{0.241pt}}
\put(1084.0,198.0){\rule[-0.200pt]{0.400pt}{0.723pt}}
\put(1082,203.67){\rule{0.241pt}{0.400pt}}
\multiput(1082.50,203.17)(-0.500,1.000){2}{\rule{0.120pt}{0.400pt}}
\put(1083.0,203.0){\usebox{\plotpoint}}
\put(1081,206.67){\rule{0.241pt}{0.400pt}}
\multiput(1081.50,206.17)(-0.500,1.000){2}{\rule{0.120pt}{0.400pt}}
\put(1082.0,205.0){\rule[-0.200pt]{0.400pt}{0.482pt}}
\put(1080,210.67){\rule{0.241pt}{0.400pt}}
\multiput(1080.50,210.17)(-0.500,1.000){2}{\rule{0.120pt}{0.400pt}}
\put(1081.0,208.0){\rule[-0.200pt]{0.400pt}{0.723pt}}
\put(1079,213.67){\rule{0.241pt}{0.400pt}}
\multiput(1079.50,213.17)(-0.500,1.000){2}{\rule{0.120pt}{0.400pt}}
\put(1080.0,212.0){\rule[-0.200pt]{0.400pt}{0.482pt}}
\put(1078,216.67){\rule{0.241pt}{0.400pt}}
\multiput(1078.50,216.17)(-0.500,1.000){2}{\rule{0.120pt}{0.400pt}}
\put(1079.0,215.0){\rule[-0.200pt]{0.400pt}{0.482pt}}
\put(1077,220.67){\rule{0.241pt}{0.400pt}}
\multiput(1077.50,220.17)(-0.500,1.000){2}{\rule{0.120pt}{0.400pt}}
\put(1078.0,218.0){\rule[-0.200pt]{0.400pt}{0.723pt}}
\put(1076,223.67){\rule{0.241pt}{0.400pt}}
\multiput(1076.50,223.17)(-0.500,1.000){2}{\rule{0.120pt}{0.400pt}}
\put(1077.0,222.0){\rule[-0.200pt]{0.400pt}{0.482pt}}
\put(1074.67,226){\rule{0.400pt}{0.482pt}}
\multiput(1075.17,226.00)(-1.000,1.000){2}{\rule{0.400pt}{0.241pt}}
\put(1076.0,225.0){\usebox{\plotpoint}}
\put(1074,230.67){\rule{0.241pt}{0.400pt}}
\multiput(1074.50,230.17)(-0.500,1.000){2}{\rule{0.120pt}{0.400pt}}
\put(1075.0,228.0){\rule[-0.200pt]{0.400pt}{0.723pt}}
\put(1072.67,233){\rule{0.400pt}{0.482pt}}
\multiput(1073.17,233.00)(-1.000,1.000){2}{\rule{0.400pt}{0.241pt}}
\put(1074.0,232.0){\usebox{\plotpoint}}
\put(1072,237.67){\rule{0.241pt}{0.400pt}}
\multiput(1072.50,237.17)(-0.500,1.000){2}{\rule{0.120pt}{0.400pt}}
\put(1073.0,235.0){\rule[-0.200pt]{0.400pt}{0.723pt}}
\put(1070.67,240){\rule{0.400pt}{0.482pt}}
\multiput(1071.17,240.00)(-1.000,1.000){2}{\rule{0.400pt}{0.241pt}}
\put(1072.0,239.0){\usebox{\plotpoint}}
\put(1070,244.67){\rule{0.241pt}{0.400pt}}
\multiput(1070.50,244.17)(-0.500,1.000){2}{\rule{0.120pt}{0.400pt}}
\put(1071.0,242.0){\rule[-0.200pt]{0.400pt}{0.723pt}}
\put(1068.67,247){\rule{0.400pt}{0.482pt}}
\multiput(1069.17,247.00)(-1.000,1.000){2}{\rule{0.400pt}{0.241pt}}
\put(1070.0,246.0){\usebox{\plotpoint}}
\put(1068,251.67){\rule{0.241pt}{0.400pt}}
\multiput(1068.50,251.17)(-0.500,1.000){2}{\rule{0.120pt}{0.400pt}}
\put(1069.0,249.0){\rule[-0.200pt]{0.400pt}{0.723pt}}
\put(1066.67,254){\rule{0.400pt}{0.482pt}}
\multiput(1067.17,254.00)(-1.000,1.000){2}{\rule{0.400pt}{0.241pt}}
\put(1068.0,253.0){\usebox{\plotpoint}}
\put(1065.67,257){\rule{0.400pt}{0.482pt}}
\multiput(1066.17,257.00)(-1.000,1.000){2}{\rule{0.400pt}{0.241pt}}
\put(1067.0,256.0){\usebox{\plotpoint}}
\put(1064.67,261){\rule{0.400pt}{0.482pt}}
\multiput(1065.17,261.00)(-1.000,1.000){2}{\rule{0.400pt}{0.241pt}}
\put(1066.0,259.0){\rule[-0.200pt]{0.400pt}{0.482pt}}
\put(1063.67,264){\rule{0.400pt}{0.482pt}}
\multiput(1064.17,264.00)(-1.000,1.000){2}{\rule{0.400pt}{0.241pt}}
\put(1065.0,263.0){\usebox{\plotpoint}}
\put(1062.67,268){\rule{0.400pt}{0.482pt}}
\multiput(1063.17,268.00)(-1.000,1.000){2}{\rule{0.400pt}{0.241pt}}
\put(1064.0,266.0){\rule[-0.200pt]{0.400pt}{0.482pt}}
\put(1061.67,271){\rule{0.400pt}{0.482pt}}
\multiput(1062.17,271.00)(-1.000,1.000){2}{\rule{0.400pt}{0.241pt}}
\put(1063.0,270.0){\usebox{\plotpoint}}
\put(1061,273.67){\rule{0.241pt}{0.400pt}}
\multiput(1061.50,273.17)(-0.500,1.000){2}{\rule{0.120pt}{0.400pt}}
\put(1062.0,273.0){\usebox{\plotpoint}}
\put(1059.67,278){\rule{0.400pt}{0.482pt}}
\multiput(1060.17,278.00)(-1.000,1.000){2}{\rule{0.400pt}{0.241pt}}
\put(1061.0,275.0){\rule[-0.200pt]{0.400pt}{0.723pt}}
\put(1059,280.67){\rule{0.241pt}{0.400pt}}
\multiput(1059.50,280.17)(-0.500,1.000){2}{\rule{0.120pt}{0.400pt}}
\put(1060.0,280.0){\usebox{\plotpoint}}
\put(1057.67,285){\rule{0.400pt}{0.482pt}}
\multiput(1058.17,285.00)(-1.000,1.000){2}{\rule{0.400pt}{0.241pt}}
\put(1059.0,282.0){\rule[-0.200pt]{0.400pt}{0.723pt}}
\put(1057,287.67){\rule{0.241pt}{0.400pt}}
\multiput(1057.50,287.17)(-0.500,1.000){2}{\rule{0.120pt}{0.400pt}}
\put(1058.0,287.0){\usebox{\plotpoint}}
\put(1056,290.67){\rule{0.241pt}{0.400pt}}
\multiput(1056.50,290.17)(-0.500,1.000){2}{\rule{0.120pt}{0.400pt}}
\put(1057.0,289.0){\rule[-0.200pt]{0.400pt}{0.482pt}}
\put(1055,294.67){\rule{0.241pt}{0.400pt}}
\multiput(1055.50,294.17)(-0.500,1.000){2}{\rule{0.120pt}{0.400pt}}
\put(1056.0,292.0){\rule[-0.200pt]{0.400pt}{0.723pt}}
\put(1054,297.67){\rule{0.241pt}{0.400pt}}
\multiput(1054.50,297.17)(-0.500,1.000){2}{\rule{0.120pt}{0.400pt}}
\put(1055.0,296.0){\rule[-0.200pt]{0.400pt}{0.482pt}}
\put(1053,300.67){\rule{0.241pt}{0.400pt}}
\multiput(1053.50,300.17)(-0.500,1.000){2}{\rule{0.120pt}{0.400pt}}
\put(1054.0,299.0){\rule[-0.200pt]{0.400pt}{0.482pt}}
\put(1052,304.67){\rule{0.241pt}{0.400pt}}
\multiput(1052.50,304.17)(-0.500,1.000){2}{\rule{0.120pt}{0.400pt}}
\put(1053.0,302.0){\rule[-0.200pt]{0.400pt}{0.723pt}}
\put(1051,307.67){\rule{0.241pt}{0.400pt}}
\multiput(1051.50,307.17)(-0.500,1.000){2}{\rule{0.120pt}{0.400pt}}
\put(1052.0,306.0){\rule[-0.200pt]{0.400pt}{0.482pt}}
\put(1049.67,310){\rule{0.400pt}{0.482pt}}
\multiput(1050.17,310.00)(-1.000,1.000){2}{\rule{0.400pt}{0.241pt}}
\put(1051.0,309.0){\usebox{\plotpoint}}
\put(1048.67,313){\rule{0.400pt}{0.482pt}}
\multiput(1049.17,313.00)(-1.000,1.000){2}{\rule{0.400pt}{0.241pt}}
\put(1050.0,312.0){\usebox{\plotpoint}}
\put(1047.67,317){\rule{0.400pt}{0.482pt}}
\multiput(1048.17,317.00)(-1.000,1.000){2}{\rule{0.400pt}{0.241pt}}
\put(1049.0,315.0){\rule[-0.200pt]{0.400pt}{0.482pt}}
\put(1046.67,320){\rule{0.400pt}{0.482pt}}
\multiput(1047.17,320.00)(-1.000,1.000){2}{\rule{0.400pt}{0.241pt}}
\put(1048.0,319.0){\usebox{\plotpoint}}
\put(1045.67,323){\rule{0.400pt}{0.482pt}}
\multiput(1046.17,323.00)(-1.000,1.000){2}{\rule{0.400pt}{0.241pt}}
\put(1047.0,322.0){\usebox{\plotpoint}}
\put(1045,325.67){\rule{0.241pt}{0.400pt}}
\multiput(1045.50,325.17)(-0.500,1.000){2}{\rule{0.120pt}{0.400pt}}
\put(1046.0,325.0){\usebox{\plotpoint}}
\put(1044,328.67){\rule{0.241pt}{0.400pt}}
\multiput(1044.50,328.17)(-0.500,1.000){2}{\rule{0.120pt}{0.400pt}}
\put(1045.0,327.0){\rule[-0.200pt]{0.400pt}{0.482pt}}
\put(1043,332.67){\rule{0.241pt}{0.400pt}}
\multiput(1043.50,332.17)(-0.500,1.000){2}{\rule{0.120pt}{0.400pt}}
\put(1044.0,330.0){\rule[-0.200pt]{0.400pt}{0.723pt}}
\put(1042,335.67){\rule{0.241pt}{0.400pt}}
\multiput(1042.50,335.17)(-0.500,1.000){2}{\rule{0.120pt}{0.400pt}}
\put(1043.0,334.0){\rule[-0.200pt]{0.400pt}{0.482pt}}
\put(1041,338.67){\rule{0.241pt}{0.400pt}}
\multiput(1041.50,338.17)(-0.500,1.000){2}{\rule{0.120pt}{0.400pt}}
\put(1042.0,337.0){\rule[-0.200pt]{0.400pt}{0.482pt}}
\put(1039.67,341){\rule{0.400pt}{0.482pt}}
\multiput(1040.17,341.00)(-1.000,1.000){2}{\rule{0.400pt}{0.241pt}}
\put(1041.0,340.0){\usebox{\plotpoint}}
\put(1038.67,344){\rule{0.400pt}{0.482pt}}
\multiput(1039.17,344.00)(-1.000,1.000){2}{\rule{0.400pt}{0.241pt}}
\put(1040.0,343.0){\usebox{\plotpoint}}
\put(1038,346.67){\rule{0.241pt}{0.400pt}}
\multiput(1038.50,346.17)(-0.500,1.000){2}{\rule{0.120pt}{0.400pt}}
\put(1039.0,346.0){\usebox{\plotpoint}}
\put(1037,349.67){\rule{0.241pt}{0.400pt}}
\multiput(1037.50,349.17)(-0.500,1.000){2}{\rule{0.120pt}{0.400pt}}
\put(1038.0,348.0){\rule[-0.200pt]{0.400pt}{0.482pt}}
\put(1036,353.67){\rule{0.241pt}{0.400pt}}
\multiput(1036.50,353.17)(-0.500,1.000){2}{\rule{0.120pt}{0.400pt}}
\put(1037.0,351.0){\rule[-0.200pt]{0.400pt}{0.723pt}}
\put(1035,356.67){\rule{0.241pt}{0.400pt}}
\multiput(1035.50,356.17)(-0.500,1.000){2}{\rule{0.120pt}{0.400pt}}
\put(1036.0,355.0){\rule[-0.200pt]{0.400pt}{0.482pt}}
\put(1034,359.67){\rule{0.241pt}{0.400pt}}
\multiput(1034.50,359.17)(-0.500,1.000){2}{\rule{0.120pt}{0.400pt}}
\put(1035.0,358.0){\rule[-0.200pt]{0.400pt}{0.482pt}}
\put(1032.67,362){\rule{0.400pt}{0.482pt}}
\multiput(1033.17,362.00)(-1.000,1.000){2}{\rule{0.400pt}{0.241pt}}
\put(1034.0,361.0){\usebox{\plotpoint}}
\put(1031.67,365){\rule{0.400pt}{0.482pt}}
\multiput(1032.17,365.00)(-1.000,1.000){2}{\rule{0.400pt}{0.241pt}}
\put(1033.0,364.0){\usebox{\plotpoint}}
\put(1031,367.67){\rule{0.241pt}{0.400pt}}
\multiput(1031.50,367.17)(-0.500,1.000){2}{\rule{0.120pt}{0.400pt}}
\put(1032.0,367.0){\usebox{\plotpoint}}
\put(1030,370.67){\rule{0.241pt}{0.400pt}}
\multiput(1030.50,370.17)(-0.500,1.000){2}{\rule{0.120pt}{0.400pt}}
\put(1031.0,369.0){\rule[-0.200pt]{0.400pt}{0.482pt}}
\put(1029,373.67){\rule{0.241pt}{0.400pt}}
\multiput(1029.50,373.17)(-0.500,1.000){2}{\rule{0.120pt}{0.400pt}}
\put(1030.0,372.0){\rule[-0.200pt]{0.400pt}{0.482pt}}
\put(1027.67,376){\rule{0.400pt}{0.482pt}}
\multiput(1028.17,376.00)(-1.000,1.000){2}{\rule{0.400pt}{0.241pt}}
\put(1026.67,378){\rule{0.400pt}{0.723pt}}
\multiput(1027.17,378.00)(-1.000,1.500){2}{\rule{0.400pt}{0.361pt}}
\put(1029.0,375.0){\usebox{\plotpoint}}
\put(1026,381.67){\rule{0.241pt}{0.400pt}}
\multiput(1026.50,381.17)(-0.500,1.000){2}{\rule{0.120pt}{0.400pt}}
\put(1027.0,381.0){\usebox{\plotpoint}}
\put(1025,384.67){\rule{0.241pt}{0.400pt}}
\multiput(1025.50,384.17)(-0.500,1.000){2}{\rule{0.120pt}{0.400pt}}
\put(1026.0,383.0){\rule[-0.200pt]{0.400pt}{0.482pt}}
\put(1024,387.67){\rule{0.241pt}{0.400pt}}
\multiput(1024.50,387.17)(-0.500,1.000){2}{\rule{0.120pt}{0.400pt}}
\put(1025.0,386.0){\rule[-0.200pt]{0.400pt}{0.482pt}}
\put(1022.67,390){\rule{0.400pt}{0.482pt}}
\multiput(1023.17,390.00)(-1.000,1.000){2}{\rule{0.400pt}{0.241pt}}
\put(1022,391.67){\rule{0.241pt}{0.400pt}}
\multiput(1022.50,391.17)(-0.500,1.000){2}{\rule{0.120pt}{0.400pt}}
\put(1024.0,389.0){\usebox{\plotpoint}}
\put(1021,394.67){\rule{0.241pt}{0.400pt}}
\multiput(1021.50,394.17)(-0.500,1.000){2}{\rule{0.120pt}{0.400pt}}
\put(1022.0,393.0){\rule[-0.200pt]{0.400pt}{0.482pt}}
\put(1019.67,397){\rule{0.400pt}{0.482pt}}
\multiput(1020.17,397.00)(-1.000,1.000){2}{\rule{0.400pt}{0.241pt}}
\put(1021.0,396.0){\usebox{\plotpoint}}
\put(1018.67,400){\rule{0.400pt}{0.482pt}}
\multiput(1019.17,400.00)(-1.000,1.000){2}{\rule{0.400pt}{0.241pt}}
\put(1020.0,399.0){\usebox{\plotpoint}}
\put(1018,402.67){\rule{0.241pt}{0.400pt}}
\multiput(1018.50,402.17)(-0.500,1.000){2}{\rule{0.120pt}{0.400pt}}
\put(1019.0,402.0){\usebox{\plotpoint}}
\put(1017,405.67){\rule{0.241pt}{0.400pt}}
\multiput(1017.50,405.17)(-0.500,1.000){2}{\rule{0.120pt}{0.400pt}}
\put(1018.0,404.0){\rule[-0.200pt]{0.400pt}{0.482pt}}
\put(1016,408.67){\rule{0.241pt}{0.400pt}}
\multiput(1016.50,408.17)(-0.500,1.000){2}{\rule{0.120pt}{0.400pt}}
\put(1017.0,407.0){\rule[-0.200pt]{0.400pt}{0.482pt}}
\put(1014.67,411){\rule{0.400pt}{0.482pt}}
\multiput(1015.17,411.00)(-1.000,1.000){2}{\rule{0.400pt}{0.241pt}}
\put(1014,412.67){\rule{0.241pt}{0.400pt}}
\multiput(1014.50,412.17)(-0.500,1.000){2}{\rule{0.120pt}{0.400pt}}
\put(1016.0,410.0){\usebox{\plotpoint}}
\put(1013,415.67){\rule{0.241pt}{0.400pt}}
\multiput(1013.50,415.17)(-0.500,1.000){2}{\rule{0.120pt}{0.400pt}}
\put(1014.0,414.0){\rule[-0.200pt]{0.400pt}{0.482pt}}
\put(1011.67,418){\rule{0.400pt}{0.482pt}}
\multiput(1012.17,418.00)(-1.000,1.000){2}{\rule{0.400pt}{0.241pt}}
\put(1013.0,417.0){\usebox{\plotpoint}}
\put(1010.67,421){\rule{0.400pt}{0.482pt}}
\multiput(1011.17,421.00)(-1.000,1.000){2}{\rule{0.400pt}{0.241pt}}
\put(1012.0,420.0){\usebox{\plotpoint}}
\put(1010,423.67){\rule{0.241pt}{0.400pt}}
\multiput(1010.50,423.17)(-0.500,1.000){2}{\rule{0.120pt}{0.400pt}}
\put(1008.67,425){\rule{0.400pt}{0.482pt}}
\multiput(1009.17,425.00)(-1.000,1.000){2}{\rule{0.400pt}{0.241pt}}
\put(1011.0,423.0){\usebox{\plotpoint}}
\put(1007.67,428){\rule{0.400pt}{0.482pt}}
\multiput(1008.17,428.00)(-1.000,1.000){2}{\rule{0.400pt}{0.241pt}}
\put(1009.0,427.0){\usebox{\plotpoint}}
\put(1007,430.67){\rule{0.241pt}{0.400pt}}
\multiput(1007.50,430.17)(-0.500,1.000){2}{\rule{0.120pt}{0.400pt}}
\put(1005.67,432){\rule{0.400pt}{0.482pt}}
\multiput(1006.17,432.00)(-1.000,1.000){2}{\rule{0.400pt}{0.241pt}}
\put(1008.0,430.0){\usebox{\plotpoint}}
\put(1004.67,435){\rule{0.400pt}{0.482pt}}
\multiput(1005.17,435.00)(-1.000,1.000){2}{\rule{0.400pt}{0.241pt}}
\put(1006.0,434.0){\usebox{\plotpoint}}
\put(1004,437.67){\rule{0.241pt}{0.400pt}}
\multiput(1004.50,437.17)(-0.500,1.000){2}{\rule{0.120pt}{0.400pt}}
\put(1005.0,437.0){\usebox{\plotpoint}}
\put(1003,440.67){\rule{0.241pt}{0.400pt}}
\multiput(1003.50,440.17)(-0.500,1.000){2}{\rule{0.120pt}{0.400pt}}
\put(1001.67,442){\rule{0.400pt}{0.482pt}}
\multiput(1002.17,442.00)(-1.000,1.000){2}{\rule{0.400pt}{0.241pt}}
\put(1004.0,439.0){\rule[-0.200pt]{0.400pt}{0.482pt}}
\put(1001,444.67){\rule{0.241pt}{0.400pt}}
\multiput(1001.50,444.17)(-0.500,1.000){2}{\rule{0.120pt}{0.400pt}}
\put(1002.0,444.0){\usebox{\plotpoint}}
\put(1000,447.67){\rule{0.241pt}{0.400pt}}
\multiput(1000.50,447.17)(-0.500,1.000){2}{\rule{0.120pt}{0.400pt}}
\put(998.67,449){\rule{0.400pt}{0.482pt}}
\multiput(999.17,449.00)(-1.000,1.000){2}{\rule{0.400pt}{0.241pt}}
\put(1001.0,446.0){\rule[-0.200pt]{0.400pt}{0.482pt}}
\put(998,451.67){\rule{0.241pt}{0.400pt}}
\multiput(998.50,451.17)(-0.500,1.000){2}{\rule{0.120pt}{0.400pt}}
\put(999.0,451.0){\usebox{\plotpoint}}
\put(997,454.67){\rule{0.241pt}{0.400pt}}
\multiput(997.50,454.17)(-0.500,1.000){2}{\rule{0.120pt}{0.400pt}}
\put(995.67,456){\rule{0.400pt}{0.482pt}}
\multiput(996.17,456.00)(-1.000,1.000){2}{\rule{0.400pt}{0.241pt}}
\put(998.0,453.0){\rule[-0.200pt]{0.400pt}{0.482pt}}
\put(995,458.67){\rule{0.241pt}{0.400pt}}
\multiput(995.50,458.17)(-0.500,1.000){2}{\rule{0.120pt}{0.400pt}}
\put(993.67,460){\rule{0.400pt}{0.482pt}}
\multiput(994.17,460.00)(-1.000,1.000){2}{\rule{0.400pt}{0.241pt}}
\put(996.0,458.0){\usebox{\plotpoint}}
\put(992.67,463){\rule{0.400pt}{0.482pt}}
\multiput(993.17,463.00)(-1.000,1.000){2}{\rule{0.400pt}{0.241pt}}
\put(992,464.67){\rule{0.241pt}{0.400pt}}
\multiput(992.50,464.17)(-0.500,1.000){2}{\rule{0.120pt}{0.400pt}}
\put(994.0,462.0){\usebox{\plotpoint}}
\put(990.67,467){\rule{0.400pt}{0.482pt}}
\multiput(991.17,467.00)(-1.000,1.000){2}{\rule{0.400pt}{0.241pt}}
\put(992.0,466.0){\usebox{\plotpoint}}
\put(989.67,470){\rule{0.400pt}{0.482pt}}
\multiput(990.17,470.00)(-1.000,1.000){2}{\rule{0.400pt}{0.241pt}}
\put(989,471.67){\rule{0.241pt}{0.400pt}}
\multiput(989.50,471.17)(-0.500,1.000){2}{\rule{0.120pt}{0.400pt}}
\put(991.0,469.0){\usebox{\plotpoint}}
\put(987.67,474){\rule{0.400pt}{0.482pt}}
\multiput(988.17,474.00)(-1.000,1.000){2}{\rule{0.400pt}{0.241pt}}
\put(987,475.67){\rule{0.241pt}{0.400pt}}
\multiput(987.50,475.17)(-0.500,1.000){2}{\rule{0.120pt}{0.400pt}}
\put(989.0,473.0){\usebox{\plotpoint}}
\put(986,478.67){\rule{0.241pt}{0.400pt}}
\multiput(986.50,478.17)(-0.500,1.000){2}{\rule{0.120pt}{0.400pt}}
\put(985,479.67){\rule{0.241pt}{0.400pt}}
\multiput(985.50,479.17)(-0.500,1.000){2}{\rule{0.120pt}{0.400pt}}
\put(987.0,477.0){\rule[-0.200pt]{0.400pt}{0.482pt}}
\put(984,482.67){\rule{0.241pt}{0.400pt}}
\multiput(984.50,482.17)(-0.500,1.000){2}{\rule{0.120pt}{0.400pt}}
\put(982.67,484){\rule{0.400pt}{0.482pt}}
\multiput(983.17,484.00)(-1.000,1.000){2}{\rule{0.400pt}{0.241pt}}
\put(985.0,481.0){\rule[-0.200pt]{0.400pt}{0.482pt}}
\put(982,486.67){\rule{0.241pt}{0.400pt}}
\multiput(982.50,486.17)(-0.500,1.000){2}{\rule{0.120pt}{0.400pt}}
\put(980.67,488){\rule{0.400pt}{0.482pt}}
\multiput(981.17,488.00)(-1.000,1.000){2}{\rule{0.400pt}{0.241pt}}
\put(983.0,486.0){\usebox{\plotpoint}}
\put(979.67,491){\rule{0.400pt}{0.482pt}}
\multiput(980.17,491.00)(-1.000,1.000){2}{\rule{0.400pt}{0.241pt}}
\put(979,492.67){\rule{0.241pt}{0.400pt}}
\multiput(979.50,492.17)(-0.500,1.000){2}{\rule{0.120pt}{0.400pt}}
\put(981.0,490.0){\usebox{\plotpoint}}
\put(977.67,495){\rule{0.400pt}{0.482pt}}
\multiput(978.17,495.00)(-1.000,1.000){2}{\rule{0.400pt}{0.241pt}}
\put(977,496.67){\rule{0.241pt}{0.400pt}}
\multiput(977.50,496.17)(-0.500,1.000){2}{\rule{0.120pt}{0.400pt}}
\put(975.67,498){\rule{0.400pt}{0.482pt}}
\multiput(976.17,498.00)(-1.000,1.000){2}{\rule{0.400pt}{0.241pt}}
\put(979.0,494.0){\usebox{\plotpoint}}
\put(975,500.67){\rule{0.241pt}{0.400pt}}
\multiput(975.50,500.17)(-0.500,1.000){2}{\rule{0.120pt}{0.400pt}}
\put(973.67,502){\rule{0.400pt}{0.482pt}}
\multiput(974.17,502.00)(-1.000,1.000){2}{\rule{0.400pt}{0.241pt}}
\put(976.0,500.0){\usebox{\plotpoint}}
\put(972.67,505){\rule{0.400pt}{0.482pt}}
\multiput(973.17,505.00)(-1.000,1.000){2}{\rule{0.400pt}{0.241pt}}
\put(972,506.67){\rule{0.241pt}{0.400pt}}
\multiput(972.50,506.17)(-0.500,1.000){2}{\rule{0.120pt}{0.400pt}}
\put(971,507.67){\rule{0.241pt}{0.400pt}}
\multiput(971.50,507.17)(-0.500,1.000){2}{\rule{0.120pt}{0.400pt}}
\put(974.0,504.0){\usebox{\plotpoint}}
\put(970,510.67){\rule{0.241pt}{0.400pt}}
\multiput(970.50,510.17)(-0.500,1.000){2}{\rule{0.120pt}{0.400pt}}
\put(968.67,512){\rule{0.400pt}{0.482pt}}
\multiput(969.17,512.00)(-1.000,1.000){2}{\rule{0.400pt}{0.241pt}}
\put(968,513.67){\rule{0.241pt}{0.400pt}}
\multiput(968.50,513.17)(-0.500,1.000){2}{\rule{0.120pt}{0.400pt}}
\put(971.0,509.0){\rule[-0.200pt]{0.400pt}{0.482pt}}
\put(966.67,516){\rule{0.400pt}{0.482pt}}
\multiput(967.17,516.00)(-1.000,1.000){2}{\rule{0.400pt}{0.241pt}}
\put(966,517.67){\rule{0.241pt}{0.400pt}}
\multiput(966.50,517.17)(-0.500,1.000){2}{\rule{0.120pt}{0.400pt}}
\put(964.67,519){\rule{0.400pt}{0.482pt}}
\multiput(965.17,519.00)(-1.000,1.000){2}{\rule{0.400pt}{0.241pt}}
\put(968.0,515.0){\usebox{\plotpoint}}
\put(964,521.67){\rule{0.241pt}{0.400pt}}
\multiput(964.50,521.17)(-0.500,1.000){2}{\rule{0.120pt}{0.400pt}}
\put(962.67,523){\rule{0.400pt}{0.482pt}}
\multiput(963.17,523.00)(-1.000,1.000){2}{\rule{0.400pt}{0.241pt}}
\put(962,524.67){\rule{0.241pt}{0.400pt}}
\multiput(962.50,524.17)(-0.500,1.000){2}{\rule{0.120pt}{0.400pt}}
\put(965.0,521.0){\usebox{\plotpoint}}
\put(961,527.67){\rule{0.241pt}{0.400pt}}
\multiput(961.50,527.17)(-0.500,1.000){2}{\rule{0.120pt}{0.400pt}}
\put(960,528.67){\rule{0.241pt}{0.400pt}}
\multiput(960.50,528.17)(-0.500,1.000){2}{\rule{0.120pt}{0.400pt}}
\put(958.67,530){\rule{0.400pt}{0.482pt}}
\multiput(959.17,530.00)(-1.000,1.000){2}{\rule{0.400pt}{0.241pt}}
\put(958,531.67){\rule{0.241pt}{0.400pt}}
\multiput(958.50,531.17)(-0.500,1.000){2}{\rule{0.120pt}{0.400pt}}
\put(962.0,526.0){\rule[-0.200pt]{0.400pt}{0.482pt}}
\put(957,534.67){\rule{0.241pt}{0.400pt}}
\multiput(957.50,534.17)(-0.500,1.000){2}{\rule{0.120pt}{0.400pt}}
\put(956,535.67){\rule{0.241pt}{0.400pt}}
\multiput(956.50,535.17)(-0.500,1.000){2}{\rule{0.120pt}{0.400pt}}
\put(954.67,537){\rule{0.400pt}{0.482pt}}
\multiput(955.17,537.00)(-1.000,1.000){2}{\rule{0.400pt}{0.241pt}}
\put(954,538.67){\rule{0.241pt}{0.400pt}}
\multiput(954.50,538.17)(-0.500,1.000){2}{\rule{0.120pt}{0.400pt}}
\put(958.0,533.0){\rule[-0.200pt]{0.400pt}{0.482pt}}
\put(953,541.67){\rule{0.241pt}{0.400pt}}
\multiput(953.50,541.17)(-0.500,1.000){2}{\rule{0.120pt}{0.400pt}}
\put(952,542.67){\rule{0.241pt}{0.400pt}}
\multiput(952.50,542.17)(-0.500,1.000){2}{\rule{0.120pt}{0.400pt}}
\put(950.67,544){\rule{0.400pt}{0.482pt}}
\multiput(951.17,544.00)(-1.000,1.000){2}{\rule{0.400pt}{0.241pt}}
\put(950,545.67){\rule{0.241pt}{0.400pt}}
\multiput(950.50,545.17)(-0.500,1.000){2}{\rule{0.120pt}{0.400pt}}
\put(948.67,547){\rule{0.400pt}{0.482pt}}
\multiput(949.17,547.00)(-1.000,1.000){2}{\rule{0.400pt}{0.241pt}}
\put(954.0,540.0){\rule[-0.200pt]{0.400pt}{0.482pt}}
\put(948,549.67){\rule{0.241pt}{0.400pt}}
\multiput(948.50,549.17)(-0.500,1.000){2}{\rule{0.120pt}{0.400pt}}
\put(946.67,551){\rule{0.400pt}{0.482pt}}
\multiput(947.17,551.00)(-1.000,1.000){2}{\rule{0.400pt}{0.241pt}}
\put(946,552.67){\rule{0.241pt}{0.400pt}}
\multiput(946.50,552.17)(-0.500,1.000){2}{\rule{0.120pt}{0.400pt}}
\put(944.67,554){\rule{0.400pt}{0.482pt}}
\multiput(945.17,554.00)(-1.000,1.000){2}{\rule{0.400pt}{0.241pt}}
\put(944,555.67){\rule{0.241pt}{0.400pt}}
\multiput(944.50,555.17)(-0.500,1.000){2}{\rule{0.120pt}{0.400pt}}
\put(943,556.67){\rule{0.241pt}{0.400pt}}
\multiput(943.50,556.17)(-0.500,1.000){2}{\rule{0.120pt}{0.400pt}}
\put(941.67,558){\rule{0.400pt}{0.482pt}}
\multiput(942.17,558.00)(-1.000,1.000){2}{\rule{0.400pt}{0.241pt}}
\put(949.0,549.0){\usebox{\plotpoint}}
\put(940.67,561){\rule{0.400pt}{0.482pt}}
\multiput(941.17,561.00)(-1.000,1.000){2}{\rule{0.400pt}{0.241pt}}
\put(940,562.67){\rule{0.241pt}{0.400pt}}
\multiput(940.50,562.17)(-0.500,1.000){2}{\rule{0.120pt}{0.400pt}}
\put(939,563.67){\rule{0.241pt}{0.400pt}}
\multiput(939.50,563.17)(-0.500,1.000){2}{\rule{0.120pt}{0.400pt}}
\put(937.67,565){\rule{0.400pt}{0.482pt}}
\multiput(938.17,565.00)(-1.000,1.000){2}{\rule{0.400pt}{0.241pt}}
\put(937,566.67){\rule{0.241pt}{0.400pt}}
\multiput(937.50,566.17)(-0.500,1.000){2}{\rule{0.120pt}{0.400pt}}
\put(935.67,568){\rule{0.400pt}{0.482pt}}
\multiput(936.17,568.00)(-1.000,1.000){2}{\rule{0.400pt}{0.241pt}}
\put(935,569.67){\rule{0.241pt}{0.400pt}}
\multiput(935.50,569.17)(-0.500,1.000){2}{\rule{0.120pt}{0.400pt}}
\put(934,570.67){\rule{0.241pt}{0.400pt}}
\multiput(934.50,570.17)(-0.500,1.000){2}{\rule{0.120pt}{0.400pt}}
\put(932.67,572){\rule{0.400pt}{0.482pt}}
\multiput(933.17,572.00)(-1.000,1.000){2}{\rule{0.400pt}{0.241pt}}
\put(932,573.67){\rule{0.241pt}{0.400pt}}
\multiput(932.50,573.17)(-0.500,1.000){2}{\rule{0.120pt}{0.400pt}}
\put(930.67,575){\rule{0.400pt}{0.482pt}}
\multiput(931.17,575.00)(-1.000,1.000){2}{\rule{0.400pt}{0.241pt}}
\put(942.0,560.0){\usebox{\plotpoint}}
\put(930,577.67){\rule{0.241pt}{0.400pt}}
\multiput(930.50,577.17)(-0.500,1.000){2}{\rule{0.120pt}{0.400pt}}
\put(928.67,579){\rule{0.400pt}{0.482pt}}
\multiput(929.17,579.00)(-1.000,1.000){2}{\rule{0.400pt}{0.241pt}}
\put(928,580.67){\rule{0.241pt}{0.400pt}}
\multiput(928.50,580.17)(-0.500,1.000){2}{\rule{0.120pt}{0.400pt}}
\put(926.67,582){\rule{0.400pt}{0.482pt}}
\multiput(927.17,582.00)(-1.000,1.000){2}{\rule{0.400pt}{0.241pt}}
\put(926,583.67){\rule{0.241pt}{0.400pt}}
\multiput(926.50,583.17)(-0.500,1.000){2}{\rule{0.120pt}{0.400pt}}
\put(925,584.67){\rule{0.241pt}{0.400pt}}
\multiput(925.50,584.17)(-0.500,1.000){2}{\rule{0.120pt}{0.400pt}}
\put(923.67,586){\rule{0.400pt}{0.482pt}}
\multiput(924.17,586.00)(-1.000,1.000){2}{\rule{0.400pt}{0.241pt}}
\put(923,587.67){\rule{0.241pt}{0.400pt}}
\multiput(923.50,587.17)(-0.500,1.000){2}{\rule{0.120pt}{0.400pt}}
\put(921.67,589){\rule{0.400pt}{0.482pt}}
\multiput(922.17,589.00)(-1.000,1.000){2}{\rule{0.400pt}{0.241pt}}
\put(921,590.67){\rule{0.241pt}{0.400pt}}
\multiput(921.50,590.17)(-0.500,1.000){2}{\rule{0.120pt}{0.400pt}}
\put(920,591.67){\rule{0.241pt}{0.400pt}}
\multiput(920.50,591.17)(-0.500,1.000){2}{\rule{0.120pt}{0.400pt}}
\put(918.67,593){\rule{0.400pt}{0.482pt}}
\multiput(919.17,593.00)(-1.000,1.000){2}{\rule{0.400pt}{0.241pt}}
\put(918,594.67){\rule{0.241pt}{0.400pt}}
\multiput(918.50,594.17)(-0.500,1.000){2}{\rule{0.120pt}{0.400pt}}
\put(916,596.17){\rule{0.482pt}{0.400pt}}
\multiput(917.00,595.17)(-1.000,2.000){2}{\rule{0.241pt}{0.400pt}}
\put(915,597.67){\rule{0.241pt}{0.400pt}}
\multiput(915.50,597.17)(-0.500,1.000){2}{\rule{0.120pt}{0.400pt}}
\put(914,598.67){\rule{0.241pt}{0.400pt}}
\multiput(914.50,598.17)(-0.500,1.000){2}{\rule{0.120pt}{0.400pt}}
\put(912.67,600){\rule{0.400pt}{0.482pt}}
\multiput(913.17,600.00)(-1.000,1.000){2}{\rule{0.400pt}{0.241pt}}
\put(912,601.67){\rule{0.241pt}{0.400pt}}
\multiput(912.50,601.17)(-0.500,1.000){2}{\rule{0.120pt}{0.400pt}}
\put(910.67,603){\rule{0.400pt}{0.482pt}}
\multiput(911.17,603.00)(-1.000,1.000){2}{\rule{0.400pt}{0.241pt}}
\put(910,604.67){\rule{0.241pt}{0.400pt}}
\multiput(910.50,604.17)(-0.500,1.000){2}{\rule{0.120pt}{0.400pt}}
\put(909,605.67){\rule{0.241pt}{0.400pt}}
\multiput(909.50,605.17)(-0.500,1.000){2}{\rule{0.120pt}{0.400pt}}
\put(907.67,607){\rule{0.400pt}{0.482pt}}
\multiput(908.17,607.00)(-1.000,1.000){2}{\rule{0.400pt}{0.241pt}}
\put(907,608.67){\rule{0.241pt}{0.400pt}}
\multiput(907.50,608.17)(-0.500,1.000){2}{\rule{0.120pt}{0.400pt}}
\put(905.67,610){\rule{0.400pt}{0.482pt}}
\multiput(906.17,610.00)(-1.000,1.000){2}{\rule{0.400pt}{0.241pt}}
\put(904,611.67){\rule{0.482pt}{0.400pt}}
\multiput(905.00,611.17)(-1.000,1.000){2}{\rule{0.241pt}{0.400pt}}
\put(903,612.67){\rule{0.241pt}{0.400pt}}
\multiput(903.50,612.17)(-0.500,1.000){2}{\rule{0.120pt}{0.400pt}}
\put(901.67,614){\rule{0.400pt}{0.482pt}}
\multiput(902.17,614.00)(-1.000,1.000){2}{\rule{0.400pt}{0.241pt}}
\put(901,615.67){\rule{0.241pt}{0.400pt}}
\multiput(901.50,615.17)(-0.500,1.000){2}{\rule{0.120pt}{0.400pt}}
\put(899.67,617){\rule{0.400pt}{0.482pt}}
\multiput(900.17,617.00)(-1.000,1.000){2}{\rule{0.400pt}{0.241pt}}
\put(899,618.67){\rule{0.241pt}{0.400pt}}
\multiput(899.50,618.17)(-0.500,1.000){2}{\rule{0.120pt}{0.400pt}}
\put(898,619.67){\rule{0.241pt}{0.400pt}}
\multiput(898.50,619.17)(-0.500,1.000){2}{\rule{0.120pt}{0.400pt}}
\put(896,621.17){\rule{0.482pt}{0.400pt}}
\multiput(897.00,620.17)(-1.000,2.000){2}{\rule{0.241pt}{0.400pt}}
\put(895,622.67){\rule{0.241pt}{0.400pt}}
\multiput(895.50,622.17)(-0.500,1.000){2}{\rule{0.120pt}{0.400pt}}
\put(893.67,624){\rule{0.400pt}{0.482pt}}
\multiput(894.17,624.00)(-1.000,1.000){2}{\rule{0.400pt}{0.241pt}}
\put(893,625.67){\rule{0.241pt}{0.400pt}}
\multiput(893.50,625.17)(-0.500,1.000){2}{\rule{0.120pt}{0.400pt}}
\put(891,626.67){\rule{0.482pt}{0.400pt}}
\multiput(892.00,626.17)(-1.000,1.000){2}{\rule{0.241pt}{0.400pt}}
\put(889.67,628){\rule{0.400pt}{0.482pt}}
\multiput(890.17,628.00)(-1.000,1.000){2}{\rule{0.400pt}{0.241pt}}
\put(889,629.67){\rule{0.241pt}{0.400pt}}
\multiput(889.50,629.17)(-0.500,1.000){2}{\rule{0.120pt}{0.400pt}}
\put(887.67,631){\rule{0.400pt}{0.482pt}}
\multiput(888.17,631.00)(-1.000,1.000){2}{\rule{0.400pt}{0.241pt}}
\put(886,632.67){\rule{0.482pt}{0.400pt}}
\multiput(887.00,632.17)(-1.000,1.000){2}{\rule{0.241pt}{0.400pt}}
\put(885,633.67){\rule{0.241pt}{0.400pt}}
\multiput(885.50,633.17)(-0.500,1.000){2}{\rule{0.120pt}{0.400pt}}
\put(883.67,635){\rule{0.400pt}{0.482pt}}
\multiput(884.17,635.00)(-1.000,1.000){2}{\rule{0.400pt}{0.241pt}}
\put(883,636.67){\rule{0.241pt}{0.400pt}}
\multiput(883.50,636.17)(-0.500,1.000){2}{\rule{0.120pt}{0.400pt}}
\put(881,638.17){\rule{0.482pt}{0.400pt}}
\multiput(882.00,637.17)(-1.000,2.000){2}{\rule{0.241pt}{0.400pt}}
\put(880,639.67){\rule{0.241pt}{0.400pt}}
\multiput(880.50,639.17)(-0.500,1.000){2}{\rule{0.120pt}{0.400pt}}
\put(879,640.67){\rule{0.241pt}{0.400pt}}
\multiput(879.50,640.17)(-0.500,1.000){2}{\rule{0.120pt}{0.400pt}}
\put(877,642.17){\rule{0.482pt}{0.400pt}}
\multiput(878.00,641.17)(-1.000,2.000){2}{\rule{0.241pt}{0.400pt}}
\put(876,643.67){\rule{0.241pt}{0.400pt}}
\multiput(876.50,643.17)(-0.500,1.000){2}{\rule{0.120pt}{0.400pt}}
\put(874,645.17){\rule{0.482pt}{0.400pt}}
\multiput(875.00,644.17)(-1.000,2.000){2}{\rule{0.241pt}{0.400pt}}
\put(873,646.67){\rule{0.241pt}{0.400pt}}
\multiput(873.50,646.17)(-0.500,1.000){2}{\rule{0.120pt}{0.400pt}}
\put(872,647.67){\rule{0.241pt}{0.400pt}}
\multiput(872.50,647.17)(-0.500,1.000){2}{\rule{0.120pt}{0.400pt}}
\put(870,649.17){\rule{0.482pt}{0.400pt}}
\multiput(871.00,648.17)(-1.000,2.000){2}{\rule{0.241pt}{0.400pt}}
\put(869,650.67){\rule{0.241pt}{0.400pt}}
\multiput(869.50,650.17)(-0.500,1.000){2}{\rule{0.120pt}{0.400pt}}
\put(867,652.17){\rule{0.482pt}{0.400pt}}
\multiput(868.00,651.17)(-1.000,2.000){2}{\rule{0.241pt}{0.400pt}}
\put(866,653.67){\rule{0.241pt}{0.400pt}}
\multiput(866.50,653.17)(-0.500,1.000){2}{\rule{0.120pt}{0.400pt}}
\put(864,654.67){\rule{0.482pt}{0.400pt}}
\multiput(865.00,654.17)(-1.000,1.000){2}{\rule{0.241pt}{0.400pt}}
\put(862.67,656){\rule{0.400pt}{0.482pt}}
\multiput(863.17,656.00)(-1.000,1.000){2}{\rule{0.400pt}{0.241pt}}
\put(861,657.67){\rule{0.482pt}{0.400pt}}
\multiput(862.00,657.17)(-1.000,1.000){2}{\rule{0.241pt}{0.400pt}}
\put(859.67,659){\rule{0.400pt}{0.482pt}}
\multiput(860.17,659.00)(-1.000,1.000){2}{\rule{0.400pt}{0.241pt}}
\put(858,660.67){\rule{0.482pt}{0.400pt}}
\multiput(859.00,660.17)(-1.000,1.000){2}{\rule{0.241pt}{0.400pt}}
\put(857,661.67){\rule{0.241pt}{0.400pt}}
\multiput(857.50,661.17)(-0.500,1.000){2}{\rule{0.120pt}{0.400pt}}
\put(855,663.17){\rule{0.482pt}{0.400pt}}
\multiput(856.00,662.17)(-1.000,2.000){2}{\rule{0.241pt}{0.400pt}}
\put(854,664.67){\rule{0.241pt}{0.400pt}}
\multiput(854.50,664.17)(-0.500,1.000){2}{\rule{0.120pt}{0.400pt}}
\put(852,666.17){\rule{0.482pt}{0.400pt}}
\multiput(853.00,665.17)(-1.000,2.000){2}{\rule{0.241pt}{0.400pt}}
\put(851,667.67){\rule{0.241pt}{0.400pt}}
\multiput(851.50,667.17)(-0.500,1.000){2}{\rule{0.120pt}{0.400pt}}
\put(849,668.67){\rule{0.482pt}{0.400pt}}
\multiput(850.00,668.17)(-1.000,1.000){2}{\rule{0.241pt}{0.400pt}}
\put(847,670.17){\rule{0.482pt}{0.400pt}}
\multiput(848.00,669.17)(-1.000,2.000){2}{\rule{0.241pt}{0.400pt}}
\put(846,671.67){\rule{0.241pt}{0.400pt}}
\multiput(846.50,671.17)(-0.500,1.000){2}{\rule{0.120pt}{0.400pt}}
\put(844,673.17){\rule{0.482pt}{0.400pt}}
\multiput(845.00,672.17)(-1.000,2.000){2}{\rule{0.241pt}{0.400pt}}
\put(842,674.67){\rule{0.482pt}{0.400pt}}
\multiput(843.00,674.17)(-1.000,1.000){2}{\rule{0.241pt}{0.400pt}}
\put(841,675.67){\rule{0.241pt}{0.400pt}}
\multiput(841.50,675.17)(-0.500,1.000){2}{\rule{0.120pt}{0.400pt}}
\put(839,677.17){\rule{0.482pt}{0.400pt}}
\multiput(840.00,676.17)(-1.000,2.000){2}{\rule{0.241pt}{0.400pt}}
\put(837,678.67){\rule{0.482pt}{0.400pt}}
\multiput(838.00,678.17)(-1.000,1.000){2}{\rule{0.241pt}{0.400pt}}
\put(835,680.17){\rule{0.482pt}{0.400pt}}
\multiput(836.00,679.17)(-1.000,2.000){2}{\rule{0.241pt}{0.400pt}}
\put(833,681.67){\rule{0.482pt}{0.400pt}}
\multiput(834.00,681.17)(-1.000,1.000){2}{\rule{0.241pt}{0.400pt}}
\put(832,682.67){\rule{0.241pt}{0.400pt}}
\multiput(832.50,682.17)(-0.500,1.000){2}{\rule{0.120pt}{0.400pt}}
\put(830,684.17){\rule{0.482pt}{0.400pt}}
\multiput(831.00,683.17)(-1.000,2.000){2}{\rule{0.241pt}{0.400pt}}
\put(828,685.67){\rule{0.482pt}{0.400pt}}
\multiput(829.00,685.17)(-1.000,1.000){2}{\rule{0.241pt}{0.400pt}}
\put(826,687.17){\rule{0.482pt}{0.400pt}}
\multiput(827.00,686.17)(-1.000,2.000){2}{\rule{0.241pt}{0.400pt}}
\put(824,688.67){\rule{0.482pt}{0.400pt}}
\multiput(825.00,688.17)(-1.000,1.000){2}{\rule{0.241pt}{0.400pt}}
\put(822,689.67){\rule{0.482pt}{0.400pt}}
\multiput(823.00,689.17)(-1.000,1.000){2}{\rule{0.241pt}{0.400pt}}
\put(820,691.17){\rule{0.482pt}{0.400pt}}
\multiput(821.00,690.17)(-1.000,2.000){2}{\rule{0.241pt}{0.400pt}}
\put(818,692.67){\rule{0.482pt}{0.400pt}}
\multiput(819.00,692.17)(-1.000,1.000){2}{\rule{0.241pt}{0.400pt}}
\put(816,694.17){\rule{0.482pt}{0.400pt}}
\multiput(817.00,693.17)(-1.000,2.000){2}{\rule{0.241pt}{0.400pt}}
\put(814,695.67){\rule{0.482pt}{0.400pt}}
\multiput(815.00,695.17)(-1.000,1.000){2}{\rule{0.241pt}{0.400pt}}
\put(812,696.67){\rule{0.482pt}{0.400pt}}
\multiput(813.00,696.17)(-1.000,1.000){2}{\rule{0.241pt}{0.400pt}}
\put(810,698.17){\rule{0.482pt}{0.400pt}}
\multiput(811.00,697.17)(-1.000,2.000){2}{\rule{0.241pt}{0.400pt}}
\put(808,699.67){\rule{0.482pt}{0.400pt}}
\multiput(809.00,699.17)(-1.000,1.000){2}{\rule{0.241pt}{0.400pt}}
\put(806,701.17){\rule{0.482pt}{0.400pt}}
\multiput(807.00,700.17)(-1.000,2.000){2}{\rule{0.241pt}{0.400pt}}
\put(803,702.67){\rule{0.723pt}{0.400pt}}
\multiput(804.50,702.17)(-1.500,1.000){2}{\rule{0.361pt}{0.400pt}}
\put(801,703.67){\rule{0.482pt}{0.400pt}}
\multiput(802.00,703.17)(-1.000,1.000){2}{\rule{0.241pt}{0.400pt}}
\put(799,705.17){\rule{0.482pt}{0.400pt}}
\multiput(800.00,704.17)(-1.000,2.000){2}{\rule{0.241pt}{0.400pt}}
\put(797,706.67){\rule{0.482pt}{0.400pt}}
\multiput(798.00,706.17)(-1.000,1.000){2}{\rule{0.241pt}{0.400pt}}
\put(794,708.17){\rule{0.700pt}{0.400pt}}
\multiput(795.55,707.17)(-1.547,2.000){2}{\rule{0.350pt}{0.400pt}}
\put(792,709.67){\rule{0.482pt}{0.400pt}}
\multiput(793.00,709.17)(-1.000,1.000){2}{\rule{0.241pt}{0.400pt}}
\put(790,710.67){\rule{0.482pt}{0.400pt}}
\multiput(791.00,710.17)(-1.000,1.000){2}{\rule{0.241pt}{0.400pt}}
\put(787,712.17){\rule{0.700pt}{0.400pt}}
\multiput(788.55,711.17)(-1.547,2.000){2}{\rule{0.350pt}{0.400pt}}
\put(785,713.67){\rule{0.482pt}{0.400pt}}
\multiput(786.00,713.17)(-1.000,1.000){2}{\rule{0.241pt}{0.400pt}}
\put(782,715.17){\rule{0.700pt}{0.400pt}}
\multiput(783.55,714.17)(-1.547,2.000){2}{\rule{0.350pt}{0.400pt}}
\put(780,716.67){\rule{0.482pt}{0.400pt}}
\multiput(781.00,716.17)(-1.000,1.000){2}{\rule{0.241pt}{0.400pt}}
\put(777,717.67){\rule{0.723pt}{0.400pt}}
\multiput(778.50,717.17)(-1.500,1.000){2}{\rule{0.361pt}{0.400pt}}
\put(774,719.17){\rule{0.700pt}{0.400pt}}
\multiput(775.55,718.17)(-1.547,2.000){2}{\rule{0.350pt}{0.400pt}}
\put(772,720.67){\rule{0.482pt}{0.400pt}}
\multiput(773.00,720.17)(-1.000,1.000){2}{\rule{0.241pt}{0.400pt}}
\put(769,722.17){\rule{0.700pt}{0.400pt}}
\multiput(770.55,721.17)(-1.547,2.000){2}{\rule{0.350pt}{0.400pt}}
\put(766,723.67){\rule{0.723pt}{0.400pt}}
\multiput(767.50,723.17)(-1.500,1.000){2}{\rule{0.361pt}{0.400pt}}
\put(763,724.67){\rule{0.723pt}{0.400pt}}
\multiput(764.50,724.17)(-1.500,1.000){2}{\rule{0.361pt}{0.400pt}}
\put(760,726.17){\rule{0.700pt}{0.400pt}}
\multiput(761.55,725.17)(-1.547,2.000){2}{\rule{0.350pt}{0.400pt}}
\put(757,727.67){\rule{0.723pt}{0.400pt}}
\multiput(758.50,727.17)(-1.500,1.000){2}{\rule{0.361pt}{0.400pt}}
\put(754,729.17){\rule{0.700pt}{0.400pt}}
\multiput(755.55,728.17)(-1.547,2.000){2}{\rule{0.350pt}{0.400pt}}
\put(751,730.67){\rule{0.723pt}{0.400pt}}
\multiput(752.50,730.17)(-1.500,1.000){2}{\rule{0.361pt}{0.400pt}}
\put(748,731.67){\rule{0.723pt}{0.400pt}}
\multiput(749.50,731.17)(-1.500,1.000){2}{\rule{0.361pt}{0.400pt}}
\put(745,733.17){\rule{0.700pt}{0.400pt}}
\multiput(746.55,732.17)(-1.547,2.000){2}{\rule{0.350pt}{0.400pt}}
\put(742,734.67){\rule{0.723pt}{0.400pt}}
\multiput(743.50,734.17)(-1.500,1.000){2}{\rule{0.361pt}{0.400pt}}
\put(738,736.17){\rule{0.900pt}{0.400pt}}
\multiput(740.13,735.17)(-2.132,2.000){2}{\rule{0.450pt}{0.400pt}}
\put(735,737.67){\rule{0.723pt}{0.400pt}}
\multiput(736.50,737.17)(-1.500,1.000){2}{\rule{0.361pt}{0.400pt}}
\put(731,738.67){\rule{0.964pt}{0.400pt}}
\multiput(733.00,738.17)(-2.000,1.000){2}{\rule{0.482pt}{0.400pt}}
\put(728,740.17){\rule{0.700pt}{0.400pt}}
\multiput(729.55,739.17)(-1.547,2.000){2}{\rule{0.350pt}{0.400pt}}
\put(724,741.67){\rule{0.964pt}{0.400pt}}
\multiput(726.00,741.17)(-2.000,1.000){2}{\rule{0.482pt}{0.400pt}}
\put(720,743.17){\rule{0.900pt}{0.400pt}}
\multiput(722.13,742.17)(-2.132,2.000){2}{\rule{0.450pt}{0.400pt}}
\put(716,744.67){\rule{0.964pt}{0.400pt}}
\multiput(718.00,744.17)(-2.000,1.000){2}{\rule{0.482pt}{0.400pt}}
\put(713,746.17){\rule{0.700pt}{0.400pt}}
\multiput(714.55,745.17)(-1.547,2.000){2}{\rule{0.350pt}{0.400pt}}
\put(708,747.67){\rule{1.204pt}{0.400pt}}
\multiput(710.50,747.17)(-2.500,1.000){2}{\rule{0.602pt}{0.400pt}}
\put(704,748.67){\rule{0.964pt}{0.400pt}}
\multiput(706.00,748.17)(-2.000,1.000){2}{\rule{0.482pt}{0.400pt}}
\put(700,750.17){\rule{0.900pt}{0.400pt}}
\multiput(702.13,749.17)(-2.132,2.000){2}{\rule{0.450pt}{0.400pt}}
\put(696,751.67){\rule{0.964pt}{0.400pt}}
\multiput(698.00,751.17)(-2.000,1.000){2}{\rule{0.482pt}{0.400pt}}
\put(691,753.17){\rule{1.100pt}{0.400pt}}
\multiput(693.72,752.17)(-2.717,2.000){2}{\rule{0.550pt}{0.400pt}}
\put(686,754.67){\rule{1.204pt}{0.400pt}}
\multiput(688.50,754.17)(-2.500,1.000){2}{\rule{0.602pt}{0.400pt}}
\put(681,755.67){\rule{1.204pt}{0.400pt}}
\multiput(683.50,755.17)(-2.500,1.000){2}{\rule{0.602pt}{0.400pt}}
\put(676,757.17){\rule{1.100pt}{0.400pt}}
\multiput(678.72,756.17)(-2.717,2.000){2}{\rule{0.550pt}{0.400pt}}
\put(671,758.67){\rule{1.204pt}{0.400pt}}
\multiput(673.50,758.17)(-2.500,1.000){2}{\rule{0.602pt}{0.400pt}}
\put(666,760.17){\rule{1.100pt}{0.400pt}}
\multiput(668.72,759.17)(-2.717,2.000){2}{\rule{0.550pt}{0.400pt}}
\put(660,761.67){\rule{1.445pt}{0.400pt}}
\multiput(663.00,761.17)(-3.000,1.000){2}{\rule{0.723pt}{0.400pt}}
\put(655,762.67){\rule{1.204pt}{0.400pt}}
\multiput(657.50,762.17)(-2.500,1.000){2}{\rule{0.602pt}{0.400pt}}
\put(649,764.17){\rule{1.300pt}{0.400pt}}
\multiput(652.30,763.17)(-3.302,2.000){2}{\rule{0.650pt}{0.400pt}}
\put(642,765.67){\rule{1.686pt}{0.400pt}}
\multiput(645.50,765.17)(-3.500,1.000){2}{\rule{0.843pt}{0.400pt}}
\put(636,767.17){\rule{1.300pt}{0.400pt}}
\multiput(639.30,766.17)(-3.302,2.000){2}{\rule{0.650pt}{0.400pt}}
\put(629,768.67){\rule{1.686pt}{0.400pt}}
\multiput(632.50,768.17)(-3.500,1.000){2}{\rule{0.843pt}{0.400pt}}
\put(622,769.67){\rule{1.686pt}{0.400pt}}
\multiput(625.50,769.17)(-3.500,1.000){2}{\rule{0.843pt}{0.400pt}}
\put(614,771.17){\rule{1.700pt}{0.400pt}}
\multiput(618.47,770.17)(-4.472,2.000){2}{\rule{0.850pt}{0.400pt}}
\put(607,772.67){\rule{1.686pt}{0.400pt}}
\multiput(610.50,772.17)(-3.500,1.000){2}{\rule{0.843pt}{0.400pt}}
\put(598,774.17){\rule{1.900pt}{0.400pt}}
\multiput(603.06,773.17)(-5.056,2.000){2}{\rule{0.950pt}{0.400pt}}
\put(589,775.67){\rule{2.168pt}{0.400pt}}
\multiput(593.50,775.17)(-4.500,1.000){2}{\rule{1.084pt}{0.400pt}}
\put(580,776.67){\rule{2.168pt}{0.400pt}}
\multiput(584.50,776.17)(-4.500,1.000){2}{\rule{1.084pt}{0.400pt}}
\put(570,778.17){\rule{2.100pt}{0.400pt}}
\multiput(575.64,777.17)(-5.641,2.000){2}{\rule{1.050pt}{0.400pt}}
\put(560,779.67){\rule{2.409pt}{0.400pt}}
\multiput(565.00,779.17)(-5.000,1.000){2}{\rule{1.204pt}{0.400pt}}
\put(548,781.17){\rule{2.500pt}{0.400pt}}
\multiput(554.81,780.17)(-6.811,2.000){2}{\rule{1.250pt}{0.400pt}}
\put(536,782.67){\rule{2.891pt}{0.400pt}}
\multiput(542.00,782.17)(-6.000,1.000){2}{\rule{1.445pt}{0.400pt}}
\put(522,783.67){\rule{3.373pt}{0.400pt}}
\multiput(529.00,783.17)(-7.000,1.000){2}{\rule{1.686pt}{0.400pt}}
\put(508,785.17){\rule{2.900pt}{0.400pt}}
\multiput(515.98,784.17)(-7.981,2.000){2}{\rule{1.450pt}{0.400pt}}
\put(491,786.67){\rule{4.095pt}{0.400pt}}
\multiput(499.50,786.17)(-8.500,1.000){2}{\rule{2.048pt}{0.400pt}}
\put(473,788.17){\rule{3.700pt}{0.400pt}}
\multiput(483.32,787.17)(-10.320,2.000){2}{\rule{1.850pt}{0.400pt}}
\put(452,789.67){\rule{5.059pt}{0.400pt}}
\multiput(462.50,789.17)(-10.500,1.000){2}{\rule{2.529pt}{0.400pt}}
\put(428,790.67){\rule{5.782pt}{0.400pt}}
\multiput(440.00,790.17)(-12.000,1.000){2}{\rule{2.891pt}{0.400pt}}
\put(399,792.17){\rule{5.900pt}{0.400pt}}
\multiput(415.75,791.17)(-16.754,2.000){2}{\rule{2.950pt}{0.400pt}}
\put(364,793.67){\rule{8.432pt}{0.400pt}}
\multiput(381.50,793.17)(-17.500,1.000){2}{\rule{4.216pt}{0.400pt}}
\put(318,795.17){\rule{9.300pt}{0.400pt}}
\multiput(344.70,794.17)(-26.697,2.000){2}{\rule{4.650pt}{0.400pt}}
\put(251,796.67){\rule{16.140pt}{0.400pt}}
\multiput(284.50,796.17)(-33.500,1.000){2}{\rule{8.070pt}{0.400pt}}
\put(931.0,577.0){\usebox{\plotpoint}}
\put(220.0,798.0){\rule[-0.200pt]{7.468pt}{0.400pt}}
\end{picture}
\caption
{$\tan\beta$ as function of $\xi=\delta 
m^2_{h_{\sm D}}/\Lambda^2=-d_{\sm D}\frac{f_3(x)}{96\pi^2}$.}
\label{tangent-mixing}
\end{figure}

As an example, in the 
case of fundamental messengers, $d_{\sm U}=0$ and 
\begin{equation}
\label{parameter}
d_{\sm D}=\sum_{i=1}^3\left(|Y_{4i}^{(5)}|^2+3|X_{4i}^{(5)}|^2\right)
\end{equation}
The result of numerical solution of eq.~(\ref{ewb}) in this theory 
is shown in Fig.~\ref{tangent-mixing}, 
where 
we set $x=1$ in the argument of the logarithm
in eq.~(\ref{toploop})

The only bound on $\delta m^2_{h_{\sm D}}$ is related to the theoretical 
limit on the value of $d_{\sm D}$, coming from eq.~(\ref{delta}),
which gives $\delta m^2_{h_{\sm D}}/\Lambda^2<3\times 10^{-4}$. The value of
$\tan{\beta}$ strongly depends on the mixing parameter $d$ and may
actually be rather small. One can show that this conclusion survives if
higher order corrections are taken into account.

On the other hand,
the parameter $\mu$ weakly depends
on mixing and grows only slightly at large values of $\xi=\delta
m^2_{h_{\sm D}}/\Lambda^2$.

\section{Concluding remarks}
We have considered mixing between the usual matter and messengers
belonging to either the fundamental or antisymmetric complete $SU(5)$
multiplets in the Minimal Gauge Mediated Model. Limits
on the corresponding coupling constants coming from various flavor 
violating processes in the lepton and the 
quark sectors and CP violating processes 
in 
the quark sector have been found in refs.~\cite{we,b-tan}: at 
$x\gtrsim 0.1$, experimentally accessible values of most of the
messenger-matter Yukawa couplings are in the range $10^{-3}\div 1$,
which does not seem unrealistic. 

In fact, there are two types of constraints on the allowed region in the space
of 
messenger-matter Yukawa couplings.
All experimental bounds coming from the flavor physics 
limit only 
the products of
different Yukawa couplings $|Y_iY_j|^{1/2}$ but not 
$Y_i$ separately. 
On the other hand, theoretical bounds, coming from the requirement of
positivity of the scalar masses, 
correspond to spherical regions  $\sum |Y_i|^2<$~const
in the space of Yukawa couplings.

There 
are basically no experimental constraints 
coming from the $\tau\to e\gamma$, $\tau\to\mu\gamma$
and $b\to s\gamma$ decays and the corresponding mixing 
terms are limited by the
 self-consistency conditions inherent in the theory.
Branching ratios of these
decays must be about $10^2\div 10^3$ $(10^4\div 10^5)$ times smaller than 
existing
experimental bounds in the case of the fundamental (antisymmetric)
messengers.
Hence this model forbids 
flavor violating $\tau$ decays and the contribution to the $b\to
s\gamma$ decay at 
the level of the next generation experiments.  

We have seen that small value of $\tan{\beta}$ can naturally 
appear in MGMM 
with messenger-matter mixing. 
It would lead to important changes in
the low energy sparticle spectrum~\cite{b-tan}. For example, in MGMM
without mixing, NLSP is a combination of $\tilde{\tau}_{\sm R}$
and $\tilde{\tau}_{\sm L}$ appearing due to large 
mixing $\Delta m^2_{\tilde{\tau}_{\sm R}-\tilde{\tau}_{\sm
L}}=2m_\tau\mu \tan{\beta}$ (we remind that in this theory
$\mu\gtrsim 500$~GeV and $\tan{\beta}\gtrsim 50$). In the presence of 
mixing the possible small value of $\tan{\beta}$ 
reduces slepton mixing and NLSP
becomes bino. This  changes significantly the 
predictions of
this model for 
 collider experiments. This type of NLSP is suitable for explaining
the CDF event~\cite{CDF} with $e^+e^-\gamma\gamma E_{\sm T}$ final 
state.

\begin{figure}[htb]
\begin{center}
{\psfig{file=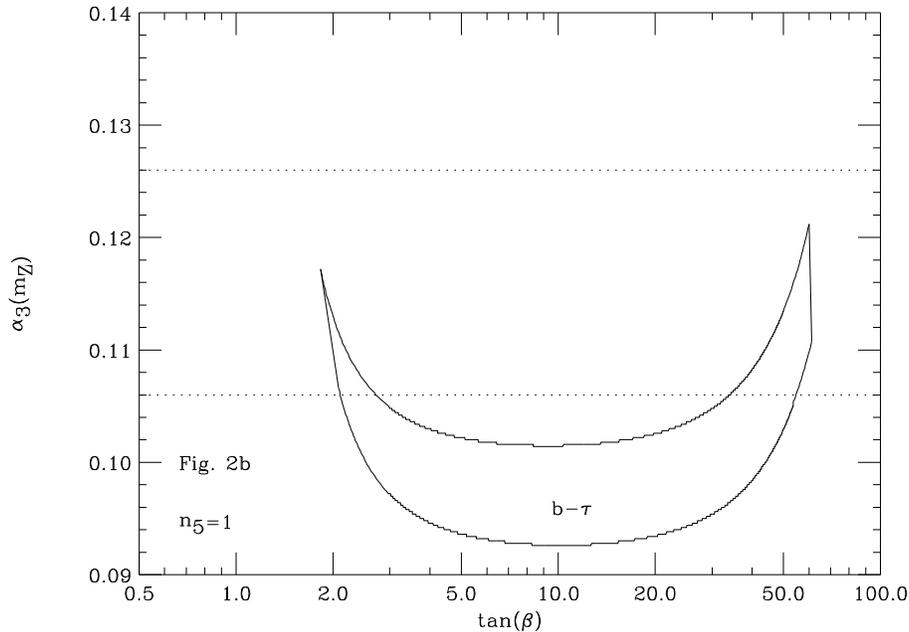,width=12cm}}
\end{center}
\caption{Constraints 
on $\tan{\beta}$ in supersymmetric $SU(5)$ model with
$n_5=1$ fundamental messenger generation 
that come from the requirement of 
$b-\tau$ unification~\cite{su5}.}
\label{b-tan-fig}
\end{figure}

One can obtain predictions
for parameters $d_{\sm U}$ and $d_{\sm D}$ 
assuming that this model is embedded into
a certain Grand Unified Theory. 
For instance, in the case of $SU(5)$ Grand
Unification with fundamental messengers one can predict
 $\tan{\beta}$ and consequently the parameter $d_{\sm D}$
from $b-\tau$ unification condition, 
$y_b(M_{\sm{GUT}})=y_\tau(M_{\sm{GUT}})$. 
Given the current uncertainties in the
determination of $\alpha_3$, two narrow
regions of $\tan\beta$ are allowed~\cite{su5} 
(see Fig.~\ref{b-tan-fig}). 
If physics below the GUT scale is indeed
described by MGMM with mixing, the predicted range
of $\tan \beta$ narrows the parameter space of
Yukawa terms and may prove 
useful for estimating the rates of flavor violation
processes.

Messenger-matter mixing can resolve the problem of fast nucleon decay in
the $SU(5)$ GUT extension of MGMM~\cite{su5}. The nucleon life-time 
 depends strongly on  $\beta$ (since $\tau (n\to
K^0\tilde{\nu})$ is proportional to $\sin^22\beta$) 
and even in the presence of
incomplete messenger multiplets it is necessary to reduce the value
of $\tan{\beta}$ in MGMM in order to suppress proton
decay~\cite{mes-split}. Messenger-matter mixing is one of the 
mechanisms which can be responsible for this reduction.

\section{Acknowledgements}
The authors are indebted V.A.Rubakov for stimulating interest and 
helpful 
suggestions. We thank F.L.Bezrukov, P.G.Tinyakov and
S.V.Troitsky for useful 
discussions. This work is supported in
part by Russian Foundation for Basic Research grant
96-02-17449a, by the  INTAS
grant 96-0457 within the research program of the
International Center for Fundamental Physics in
Moscow and by ISSEP fellowships.

%%%%%%%%%%%%%%%%%%%%%%%%%%%%%%%%%%%%%%%%%%%%%%%%%%%%%% 
\def\ijmp#1#2#3{{\it Int. Jour. Mod. Phys. }{\bf #1~}(19#2)~#3}
\def\pl#1#2#3{{\it Phys. Lett. }{\bf B#1~}(19#2)~#3}
\def\zp#1#2#3{{\it Z. Phys. }{\bf C#1~}(19#2)~#3}
\def\prl#1#2#3{{\it Phys. Rev. Lett. }{\bf #1~}(19#2)~#3}
\def\rmp#1#2#3{{\it Rev. Mod. Phys. }{\bf #1~}(19#2)~#3}
\def\prep#1#2#3{{\it Phys. Rep. }{\bf #1~}(19#2)~#3}
\def\pr#1#2#3{{\it Phys. Rev. }{\bf D#1~}(19#2)~#3}
\def\np#1#2#3{{\it Nucl. Phys. }{\bf B#1~}(19#2)~#3}
\def\mpl#1#2#3{{\it Mod. Phys. Lett. }{\bf #1~}(19#2)~#3}
\def\arnps#1#2#3{{\it Annu. Rev. Nucl. Part. Sci. }{\bf #1~}(19#2)~#3}
\def\sjnp#1#2#3{{\it Sov. J. Nucl. Phys. }{\bf #1~}(19#2)~#3}
\def\jetp#1#2#3{{\it JETP Lett. }{\bf #1~}(19#2)~#3}
\def\app#1#2#3{{\it Acta Phys. Polon. }{\bf #1~}(19#2)~#3}
\def\rnc#1#2#3{{\it Riv. Nuovo Cim. }{\bf #1~}(19#2)~#3}
\def\ap#1#2#3{{\it Ann. Phys. }{\bf #1~}(19#2)~#3}
\def\ptp#1#2#3{{\it Prog. Theor. Phys. }{\bf #1~}(19#2)~#3}
\def\spu#1#2#3{{\it Sov. Phys. Usp.}{\bf #1~}(19#2)~#3}
\addcontentsline{toc}{section}{Литература}
%%%%%%%%%%%%%%%%%%%%%%%%%%%%%%%%%%%%%%%%%%%%%%%%%%%%%%%

\end{document}